%% file: main.tex
\PassOptionsToPackage{draft}{hyperref}

\documentclass[twocolumn]{autart}

\usepackage{graphicx}
\usepackage{amsmath,amsfonts,amssymb}
\usepackage{mathtools}
\usepackage{verbatim}
\usepackage{algorithmic}
\usepackage{array}
\usepackage{textcomp}
\usepackage{stfloats}
\usepackage{url}
\usepackage{balance}
\usepackage[noadjust]{cite}
\usepackage{colortbl}
\usepackage{subcaption}
\usepackage{arydshln}
\usepackage{tabularx}
\usepackage{epstopdf}
\usepackage{tikz}
\usepackage{pgfplots}
\pgfplotsset{compat=1.18}
\usetikzlibrary{spy}
\usepackage[hidelinks]{hyperref} 

\def\BibTeX{{\rm B\kern-.05em{\sc i\kern-.025em b}\kern-.08em
    T\kern-.1667em\lower.7ex\hbox{E}\kern-.125emX}}

\newtheorem{theorem}{Theorem}
\newtheorem{remark}{Remark}
\newtheorem{definition}{Definition}
\newtheorem{lemma}{Lemma}
\newtheorem{corollary}{Corollary}

\newtheorem{assumption}{Assumption}

\begin{document}

\begin{frontmatter}

\title{On Suboptimal Safety-Critical Tracking Controller Design} 

\thanks[footnoteinfo]{}

\author[Paestum]{Yazdan Batmani}\ead{y.batmani@uok.ac.ir},    
\author[Rome]{Saber Omidi}\ead{saber.omidi@unh.edu}               

\address[Paestum]{Department of Electrical Engineering, University of Kurdistan, Sanandaj, Iran}  
\address[Rome]{Department of Mechanical Engineering,
University of New Hampshire, Durham, NH 03824, USA}             


\begin{abstract}                          
This paper proposes a novel framework for safety-critical optimal trajectory tracking in nonlinear systems based on the state-dependent Riccati equation (SDRE) methodology. By embedding barrier states into the system dynamics, the proposed strategy simultaneously ensures safety and tracking requirements, even in scenarios where these objectives may be inherently conflicting. A discounted pseudo-quadratic cost function is formulated to achieve a suboptimal trade-off between tracking accuracy, control effort, and safety objective. We present two distinct controller designs: one utilizing a single barrier state to enforce overall safety constraints, and another employing multiple barrier states to individually tuning the system's conservatism with respect to each safety constraint, providing enhanced flexibility in tuning the system’s conservatism toward individual constraints. We establish sufficient conditions to ensure the solvability of the associated Riccati equations. The proposed safe controller is well-suited for real-time implementation in practical systems, given its reasonable computational requirements and compatibility with widely available embedded microprocessors. This is supported by simulation studies involving a mechanical system and a mobile robot collision avoidance scenario, where the safe SDRE controller consistently maintained safety while achieving trajectory tracking objectives in challenging conditions. Additionally, experimental results on a cable-driven parallel robot further demonstrate the practical applicability and effectiveness of the proposed method in real-world control tasks.
\end{abstract}

\end{frontmatter}

\input{Introduction}

\input{Problem_description}

\input{Safe_SDRE}

 \input{simulation}
\input{Experiment}

\section{Conclusion}\label{conclusion}
This paper presented a novel safety-critical trajectory tracking control framework for nonlinear systems based on the SDRE methodology. By introducing well-defined barrier states into the system dynamics, the proposed approach achieves a practical balance between safety requirements and tracking performance, even in cases where these objectives may conflict. A discounted pseudo-quadratic cost function was formulated to capture the trade-offs between control effort, tracking accuracy, and safety objectives. Two distinct controller designs were developed: one employing a single barrier state to impose a global safety constraint and another utilizing multiple barrier states to independently modulate conservatism with respect to individual safety constraints. This multi-structure design offers enhanced flexibility for adjusting safety margins in complex, multi-constraint environments. Theoretical guarantees were provided by establishing sufficient conditions for the solvability of the associated Riccati equations, ensuring the stabilizability of the augmented system under the proposed framework. Comprehensive simulation studies on a mechanical system and a mobile robot collision avoidance task demonstrated the capability of the safe SDRE controllers to maintain safety while achieving accurate tracking in challenging scenarios. Furthermore, experimental validation on a cable-driven parallel robot confirmed the practical viability and effectiveness of the proposed method for real-world applications. The comparison of computational times between the proposed SSDRE controller, nonlinear MPC, and CBF-QP controllers demonstrates that the proposed method is computationally efficient and suitable for practical application in safety-critical control systems. In our future work, we explore extensions of the proposed framework to networked control systems, integration with event-triggered mechanisms to improve communication efficiency. Additionally, we aim to generalize the results of this study to develop $H_\infty$ controllers for nonlinear systems subject to external disturbances.

\bibliographystyle{plain}        
\bibliography{autosam}           



\end{document}

%% file: Introduction.tex
\section{Introduction}\label{sec: Intro}
Ensuring safety in control systems is essential to prevent hazardous events and maintain operational integrity, particularly in safety-critical applications like automotive, aerial, multi-robot, and energy systems \cite{xu2015robustness,singletary2022onboard,cheng2020safe,chen2020safety}. Since constraints are inherent to control systems, various strategies have been developed to manage them. Early methods relied on set invariance via Lyapunov analyses \cite{khalil}, while reference governors adjusted system references to maintain constraints \cite{garone2017reference}. More recently, control barrier functions (CBFs) combined with control Lyapunov functions (CLFs) in quadratic programs (QPs) have enabled multi-objective control designs for complex, nonlinear systems \cite{ames2016control,li2023survey,cohen2024safety}. However, conflicts between safety constraints and other control objectives often arise. In CLF-CBF QPs, system stability may be compromised to preserve safety by relaxing the CLF \cite{ames2016control}. An alternative proposed in \cite{almubarak2021safety} introduces barrier states integrated with system dynamics, providing a general framework for multi-objective problems. However, this can render the augmented system unstabilizable, which was addressed by incorporating an additional stabilizing term in the barrier state dynamics.

Recent advancements in safety-critical control, initially centered on stabilization, have expanded to tracking problems. Methods include adaptive control with nonlinear reference models \cite{arabi2020safety}, model-free safe controllers for robotic systems \cite{molnar2021model}, and integral control barrier functions combined with QPs for trajectory adjustments \cite{ames2020integral}. A major challenge occurs when desired references conflict with safety constraints. To handle such conflicted multi-objective cases, \cite{dong2024safety} proposed a design featuring a performance indicator function (constructed similarly to a CBF) to mediate the conflict between decreasing the tracking error and maintaining safety. Their control law combines a standard tracking controller with a safety controller component, ensuring the system remains safe while striving to reduce tracking error.

Achieving optimal performance in control design remains challenging, as solving the necessary conditions from Pontryagin's maximum principle and dynamic programming is often intractable \cite{abu2005nearly,kiumarsi2017optimal}. To address this, techniques like model predictive control (MPC) have been used for optimal control synthesis \cite{mayne2000constrained}. Inspired by linear quadratic regulators (LQRs), the state-dependent Riccati equation (SDRE) framework offers a practical approach for nonlinear systems by representing them as state-dependent linear forms \cite{batmani2016nonlinear}. This state-dependent coefficient (SDC) representation preserves system nonlinearities without approximation. SDRE is valued for its simplicity, performance tunability via weighting matrices, and robustness to uncertainties and disturbances, making it suitable for high-dimensional systems. Comprehensive surveys in \cite{ccimen2008state,ccimen2010systematic,cimen2012survey} affirm its wide applicability and effectiveness across diverse control applications. Given the importance of handling safety constraints in control design, several SDRE extensions have been proposed. In \cite{devi2021barrier}, barrier function-based weighting matrices were introduced to address symmetric state constraints. This method is unable to address broader safety specifications, including asymmetric state constraints, and is not applicable to trajectory tracking scenarios. To handle more general constraint formulations, \cite{cloutier2001state} proposed a modified cost function that includes additional terms representing constraints as functions of the system states. While effective for designing state-feedback regulators for stabilization, this method is not applicable to trajectory tracking problems.

This work addresses optimal control for nonlinear systems with safety constraints, aiming to design a safety-critical trajectory tracking controller that minimizes a pseudo-quadratic cost function of tracking error and control effort. To solve this problem, inspired from \cite{almubarak2021safety,kuperman2023improved}, we first define a general form of barrier states. We then propose two SDRE-based solutions: the first employs a single barrier state to address all safety constraints collectively, while the second introduces 
multiple dedicated barrier states, each corresponding to an individual constraint. The multi-barrier formulation offers critical advantages in control design flexibility—by independently weighting each barrier state, the controller can shape the system's conservatism with respect to specific constraints. For instance, stricter safety margins can be imposed on high-priority constraints while allowing more relaxed responses to others. Such constraint-specific tuning is impossible with a single barrier state, which treats all constraints uniformly and may lead to overly conservative behavior. By augmenting the system dynamics with either architecture, we derive SDC forms that guarantee solvability of the associated Riccati equations. The effectiveness of the proposed method, termed safe SDRE (SSDRE), is demonstrated through simulation case studies involving mechanical systems and mobile robots, highlighting its ability to prioritize safety. Furthermore, we experimentally validate the effectiveness of the proposed SSDRE controller by implementing it on a cable-driven parallel robot, demonstrating its practical applicability in real-world safety-critical systems. By comparing the computational time required by the proposed SSDRE controller with that of a nonlinear MPC and a CBF-QP controller, it can be concluded that the proposed method provides a computationally efficient solution suitable for use in real-time safety-critical control systems. The main contributions of this paper can be summarized as follows:





$\bullet$ Proposing a safety-critical SDRE-based control framework with embedded barrier states.

$\bullet$ Introducing a pseudo-quadratic cost for suboptimal trade-offs between tracking and safety.

$\bullet$ Presenting single and multiple barrier state designs for tunable conservatism.

$\bullet$ Providing theoretical solvability conditions and validating the approach via simulations and experiments.

The paper proceeds as follows. In Section~\ref{sec:problem_statement}, the problem under consideration is defined, and a brief overview of the conventional SDRE controller is provided. In Section \ref{sec:main1}, the proposed safety-critical tracking controller is presented. In Section \ref{simulation}, we investigate the effectiveness of the proposed method using two simulation case studies with different scenarios. Section~\ref{sec:experiment} reports the experimental validation of the SSDRE controller using a laboratory-scale cable-suspended planar parallel robot. Finally, in Section \ref{conclusion}, we summarize the preceding sections' key findings, contributions, and limitations.

\noindent\textbf{Notations:} The transpose of a matrix \(\boldsymbol{P}\) is denoted by \(\boldsymbol{P}^\top\). $\Vert \cdot\Vert$ shows the Euclidean norm of a vector. $\boldsymbol{I}_k$ shows the $k\times k$ identity matrix. A matrix $\boldsymbol{P}\in\mathbb{R}^{n\times n}$
is said to be positive definite (positive semi-definite), if for any nonzero vector $\boldsymbol{x}\in\mathbb{R}^n$, it satisfies $\boldsymbol{x}^\top \boldsymbol{P}\boldsymbol{x}>0$ ($\boldsymbol{x}^\top \boldsymbol{P}\boldsymbol{x}\geq 0$). The intersection and union of two sets $\Omega_1$ and $\Omega_2$ are shown with $\Omega_1\cap\Omega_2$ and $\Omega_1\cup\Omega_2$, respectively. The set of all eigenvalues of a matrix $\boldsymbol{A}$ is shown with $\lambda({\boldsymbol{A}})$. The diagonal matrix $\boldsymbol{P}\in\mathbb{R}^{n\times n}$, whose diagonal entries are $p_1, \dots, p_n$, is denoted as $\boldsymbol{P} = \mathrm{\textbf{diag}}(p_1, \dots, p_n)$. 
A function $\boldsymbol{f}: \mathbb{R}^n \rightarrow \mathbb{R}^m$ belongs to $C^k$ on a domain $\mathcal{D}$ if it possesses continuous partial derivatives of all orders up to $k$ on $\mathcal{D}$. If $k = \infty$, $\boldsymbol{f}$ is called smooth.

%% file: Problem_description.tex
\section{Problem Description and SDRE Overview}\label{sec:problem_statement}
This section begins with a description of the problem formulation, followed by an overview of the SDRE method.

\subsection{Problem Under Consideration}\label{probmel}
Consider the multi-input-multi-output nonlinear system
\begin{align}
   \dot{\boldsymbol{x}}(t)&=\boldsymbol{f}\big(\boldsymbol{x}(t)\big)+\boldsymbol{g}\big(\boldsymbol{x}(t)\big)\boldsymbol{u}(t),\label{dynamics}\\
   \boldsymbol{y}(t)&=\boldsymbol{h}\big(\boldsymbol{x}(t)\big),\label{output}
\end{align}
where $\boldsymbol{x}\!\in\! \mathbb{R}^{n}$, $\boldsymbol{u}\!\in\! \mathbb{R}^{{m}}$, and $\boldsymbol{y}\!\in\! \mathbb{R}^{{l}}$ are the system state, the control input, and the system output, respectively; $\boldsymbol{f}\!:\!\mathbb{R}^{{n}}\rightarrow \mathbb{R}^{{n}}$, $\boldsymbol{g}\!:\!\mathbb{R}^{{n}}\rightarrow \mathbb{R}^{{m}}$, and $\boldsymbol{h}\!:\!\mathbb{R}^{{n}}\rightarrow \mathbb{R}^{{l}}$ are at least $C^1$ functions. We assume $\boldsymbol{f}(\boldsymbol{0})=\boldsymbol{0}$ and $\boldsymbol{g}(\boldsymbol{x})\neq \boldsymbol{0}$ for all $\boldsymbol{x}$. The  problem is to design a controller that fulfills the following two objectives\textit{}\footnote{For brevity, we omit explicit time dependence of variables throughout this paper unless clarity requires it.}:

$\bullet$ \textbf{Safety constraint}: The overall safe set is defined as
\begin{equation}\label{safe}
\mathcal{S} \coloneqq \bigcap_{i=1}^{N} \mathcal{S}_i,
\end{equation}
and it must remain forward invariant. This implies that if the system starts from an initial state 
\(\boldsymbol{x}(0) = \boldsymbol{x}_0 \in \mathcal{S}\), the state trajectory should satisfy 
\(\boldsymbol{x}(t) \in \mathcal{S}\) for all \(t > 0\). 
Here, \(N \geq 1\) represents the total number of individual safety constraints, with each safe set defined by
\begin{equation}\label{si}
\mathcal{S}_i \coloneqq \left\{ \boldsymbol{x} \in \mathbb{R}^n \ \middle| \ s_i(\boldsymbol{x}) > 0 \right\},
\end{equation}
where \(s_i(\boldsymbol{x})\!:\! \mathbb{R}^n\! \rightarrow \!\mathbb{R}\) (\(i = 1, \dots, N\)) is a smooth function.

$\bullet$ \textbf{Optimal tracking}: The output \(\boldsymbol{y}\) must track the desired output \(\boldsymbol{y}_\mathrm{d}\) while minimizing the cost function
\begin{equation}\label{jjj}
        \mathcal{J}=\int_{0}^{\infty} \exp({-2\gamma t})\big(\boldsymbol{e}^\top \boldsymbol{Q}(\boldsymbol{x})\boldsymbol{e}+\boldsymbol{u}^\top \boldsymbol{R}(\boldsymbol{x})\boldsymbol{u}\big)dt, 
    \end{equation}
where \(\boldsymbol{e} \coloneqq \boldsymbol{y} - \boldsymbol{y}_\mathrm{d}\) is the tracking error, \(\gamma > 0\) is the discount factor, and \(\boldsymbol{Q}(\boldsymbol{x})\), \(\boldsymbol{R}(\boldsymbol{x})\) are  $C^1$ functions from \(\mathbb{R}^n\) to \(\mathbb{R}^{l \times l}\) and \(\mathbb{R}^{m \times m}\), respectively, satisfying \(\boldsymbol{Q}^\top(\boldsymbol{x}) = \boldsymbol{Q}(\boldsymbol{x}) \geq \mathbf{0}\) and \(\boldsymbol{R}^\top(\boldsymbol{x}) = \boldsymbol{R}(\boldsymbol{x}) > \mathbf{0}\) for all \(\boldsymbol{x}\in\mathcal{S}\).

In our problem, we assume that $\boldsymbol{y}_\mathrm{d}$ is the output of the following system with the initial condition $\boldsymbol{v}(0)=\boldsymbol{v}_{0}$:
\begin{align}
\dot{\boldsymbol{v}}&=\boldsymbol{f}_\mathrm{d}(\boldsymbol{v}),\label{eq20}\\
    \boldsymbol{y}_\mathrm{d}&=\boldsymbol{h}_\mathrm{d}(\boldsymbol{v}),\label{eq20a}
   \end{align}
where \(\boldsymbol{v} \in \mathbb{R}^{n_\mathrm{d}}\) and \(\boldsymbol{y}_\mathrm{d} \in \mathbb{R}^{l}\) are the state and output of the system \eqref{eq20}-\eqref{eq20a}; \(\boldsymbol{f}_\mathrm{d} : \mathbb{R}^{n_\mathrm{d}} \to \mathbb{R}^{n_\mathrm{d}}\) and \(\boldsymbol{h}_\mathrm{d} : \mathbb{R}^{n_\mathrm{d}} \to \mathbb{R}^{l}\) are at least $C^1$ functions with \(\boldsymbol{f}_\mathrm{d}(\boldsymbol{0}) = \boldsymbol{0}\) and \(\boldsymbol{h}_\mathrm{d}(\boldsymbol{0}) = \boldsymbol{0}\). 

\noindent\textbf{Optimal Safe Trajectory Tracking Problem Statement:}  
For the system \eqref{dynamics}–\eqref{output}, find \(\boldsymbol{u}\) that renders the safe set \(\mathcal{S}\) (defined in \eqref{safe}) forward invariant while minimizing the discounted cost function \eqref{jjj}.

\subsection{Conventional SDRE Control}
Consider the dynamics \eqref{dynamics} with the cost function
    \begin{equation}\label{jj}
        J=\int_{0}^{\infty} \Big(\boldsymbol{x}^\top \boldsymbol{Q}(\boldsymbol{x})\boldsymbol{x}+\boldsymbol{u}^\top \boldsymbol{R}(\boldsymbol{x})\boldsymbol{u}\Big) \,dt, 
    \end{equation}
where \(\boldsymbol{Q}(\boldsymbol{x}) : \mathbb{R}^n \to \mathbb{R}^{n \times n}\) and \(\boldsymbol{R}(\boldsymbol{x}) : \mathbb{R}^n \to \mathbb{R}^{m \times m}\) are \(C^1\) functions satisfying \(\boldsymbol{Q}^\top(\boldsymbol{x}) = \boldsymbol{Q}(\boldsymbol{x}) \geq \mathbf{0}\) and \(\boldsymbol{R}^\top(\boldsymbol{x}) = \boldsymbol{R}(\boldsymbol{x}) > \mathbf{0}\) for all \(\boldsymbol{x}\). The first step in designing an SDRE controller involves expressing \(\boldsymbol{f}(\boldsymbol{x})\) in its SDC form as $\boldsymbol{f}(\boldsymbol{x}) = \boldsymbol{A}(\boldsymbol{x}) \boldsymbol{x}$ and rewrite \eqref{dynamics} as follows:
\begin{equation}\label{sdcf}
\dot{\boldsymbol{x}} = \boldsymbol{A}(\boldsymbol{x}) \boldsymbol{x}+\boldsymbol{g}(\boldsymbol{x}) \boldsymbol{u},
\end{equation}
where \(\boldsymbol{A}(\boldsymbol{x}) \!\in\! \mathbb{R}^{n \times n}\). Provided that \(\boldsymbol{f}(\boldsymbol{0})\! =\! \boldsymbol{0}\) and \(\boldsymbol{f} \in C^1\), a continuous matrix-valued function \(\boldsymbol{A}(\boldsymbol{x})\) satisfying this factorization always exists \cite{cimen2012survey}. In the second step, the following SDRE is solved to find $\boldsymbol{P}(\boldsymbol{x})\in\mathbb{R}^{n\times n}$:
\begin{align}\label{sdre}
&\boldsymbol{A}^\top(\boldsymbol{x})\boldsymbol{P}(\boldsymbol{x})+\boldsymbol{P}(\boldsymbol{x})\boldsymbol{A}(\boldsymbol{x})-\\{}
    &\boldsymbol{P}(\boldsymbol{x})\boldsymbol{g}(\boldsymbol{x})\boldsymbol{R}^{-1}(\boldsymbol{x})\boldsymbol{g}^\top(\boldsymbol{x})\boldsymbol{P}(\boldsymbol{x})=-\boldsymbol{Q}(\boldsymbol{x}).\nonumber
\end{align}

Finally, the SDRE-based control law is derived as $\boldsymbol{u}=-\boldsymbol{K}(\boldsymbol{x})\boldsymbol{x}$ where $\boldsymbol{K}:\mathbb{R}^n\rightarrow\mathbb{R}^{m\times n}$ is as follows:
\begin{align}
    \boldsymbol{K}(\boldsymbol{x})=\boldsymbol{R}^{-1}(\boldsymbol{x})\boldsymbol{g}^\top(\boldsymbol{x})\boldsymbol{P}(\boldsymbol{x}).
\end{align}

\begin{definition}
For the nonlinear system \eqref{dynamics}-\eqref{output},  \(\big(\boldsymbol{A}(\boldsymbol{x}),\) \(\boldsymbol{g}(\boldsymbol{x})\big)\) is  pointwise stabilizable in \(\mathcal{D} \subseteq \mathbb{R}^n\) with \(\boldsymbol{0} \in \mathcal{D}\) if it is stabilizable in the linear sense for all \(\boldsymbol{x} \in \mathcal{D}\) \textup{\cite{ccimen2008state}}.
\end{definition}
\begin{definition}
For the system \eqref{dynamics}-\eqref{output}, $\big(\boldsymbol{A}(\boldsymbol{x}),\boldsymbol{C}(\boldsymbol{x})\big)$ is pointwise detectable in \(\mathcal{D} \subseteq \mathbb{R}^n\) with \(\boldsymbol{0} \in \mathcal{D}\) if $\big(\boldsymbol{A}(\boldsymbol{x}),\boldsymbol{C}(\boldsymbol{x})\big)$ is detectable in the linear sense for all $\boldsymbol{x}\in\mathcal{D}$. Here, $\boldsymbol{C}(\boldsymbol{x})$ satisfies $\boldsymbol{Q}(\boldsymbol{x})=\boldsymbol{C}^\top(\boldsymbol{x})\boldsymbol{C}(\boldsymbol{x})$ and has full rank \textup{\cite{ccimen2008state}}. 
\end{definition}

The following theorem, taken from \cite{mracek1998control}, investigates the stability of the SDRE closed-loop solution.

\begin{theorem}
Consider the system \eqref{dynamics}–\eqref{output}. Assume that in \(\mathcal{D} \subseteq \mathbb{R}^n\) with \(\boldsymbol{0} \in \mathcal{D}\), $\big(\boldsymbol{A}(\boldsymbol{x}), \boldsymbol{g}(\boldsymbol{x})\big)$ is pointwise stabilizable, $\big(\boldsymbol{A}(\boldsymbol{x}), \boldsymbol{C}(\boldsymbol{x})\big)$ is pointwise detectable, and $\boldsymbol{A}(\boldsymbol{x})$, $\boldsymbol{g}(\boldsymbol{x})$, $\boldsymbol{Q}(\boldsymbol{x})$, and $\boldsymbol{R}(\boldsymbol{x})$ are $C^1$. Then, the SDRE-based control $\boldsymbol{u}(\boldsymbol{x}) = -\boldsymbol{R}^{-1}(\boldsymbol{x}) \boldsymbol{g}^\top(\boldsymbol{x}) \boldsymbol{P}(\boldsymbol{x}) \boldsymbol{x}$, with $\boldsymbol{P}(\boldsymbol{x})$ solving \eqref{sdre}, locally asymptotically stabilizes the origin \textup{\cite{mracek1998control}}.
\end{theorem}


%% file: Safe_SDRE.tex
\section{Proposed Safe Tracking Controller}\label{sec:main1}

In this section, we extend the SDRE control framework to address the optimal safe trajectory tracking problem introduced in Section~\ref{probmel}. Section~\ref{singles} approaches this with a single barrier state, while Section~\ref{multis} refines it by introducing \(N\) barrier states, each assigned to a safety constraint. This allows individual weighting and adjustment of each constraint’s conservativeness. Both methods handle multiple safety constraints, but the key advantage of the second is its ability to independently tune the conservativeness of each constraint via its corresponding barrier state (see Section~\ref{conf}).

\subsection{Solution with One Barrier State}\label{singles}
Consider the nonlinear system described by \eqref{dynamics} and \eqref{output} with the safe set \eqref{safe}. Inspired from  \cite{almubarak2021safety}, we define the barrier state $z(t)\in\mathbb{R}$ as follows:
\begin{align}\label{smct_1}
z(t):=\frac{p\big(\boldsymbol{x}(t)\big)}{q\Big(s\big(\boldsymbol{x}(t)\big)\Big)},
\end{align}
where $p:\mathbb{R}^n\rightarrow\mathbb{R}$ and $q:\mathbb{R}\rightarrow\mathbb{R}$ are user-defined smooth functions for $\boldsymbol{x}\in\mathcal{S}$, with $p(\boldsymbol{0})=q(0)=0$ and $s\coloneqq \prod_{i=1}^N s_i$. According to the barrier state definition in \eqref{smct_1}, the safety constraints $s_i(\boldsymbol{x})>0$ for $i=1,\dots,N$ are satisfied if and only if $z$ remains bounded, i.e., $|z(t)|<\infty$ for all $t\geq 0$. Thus, the approach is to augment the dynamics of the barrier state with the system dynamics, and design a controller for the resulted augmented system to 1) ensure the safe operation of the system and 2) achieve the tracking objective.  

From \eqref{smct_1}, the dynamics of $z$ are derived as follows:
\begin{align}\label{gg}
    \dot{z}=\Big(\frac{\partial p(\boldsymbol{x})}{\partial \boldsymbol{x}}-z\frac{\partial q\big(s(\boldsymbol{x})\big)}{\partial s(\boldsymbol{x})}\frac{\partial s(\boldsymbol{x})}{\partial \boldsymbol{x}}\Big)^\top\frac{\dot{\boldsymbol{x}}}{q\big(s(\boldsymbol{x})\big)}.\nonumber \\[-1cm]
\end{align}

Using \eqref{sdcf}, \eqref{gg} can be rewritten as follows:
\begin{align}\label{gg1}
    \dot{z}\!=\!\Big(\!\frac{\partial p(\boldsymbol{x})}{\partial \boldsymbol{x}}\!-\!z\frac{\partial q\big(s(\boldsymbol{x})\big)}{\partial s(\boldsymbol{x})}\frac{\partial s(\boldsymbol{x})}{\partial \boldsymbol{x}}\Big)^\top\!\frac{\boldsymbol{A}(\boldsymbol{x})\boldsymbol{x}\!+\!\boldsymbol{g}(\boldsymbol{x})\boldsymbol{u}}{q\big(s(\boldsymbol{x})\big)}.\nonumber \\[-1cm]
\end{align}

Since we assume $p(\cdot)$ and $q(\cdot)$ are smooth functions, \eqref{gg1} can be rewritten as its SDC form 
\begin{align}\label{sdcz}
    \dot{z}=\boldsymbol{\alpha}^\top(\boldsymbol{x},z)\boldsymbol{x}+\boldsymbol{\beta}^\top(\boldsymbol{x},z)\boldsymbol{u},
\end{align}
where $\boldsymbol{\alpha}:\mathbb{R}^{n+1}\rightarrow\mathbb{R}^{n}$ and $\boldsymbol{\beta}:\mathbb{R}^{n+1}\rightarrow\mathbb{R}^{m}$. Note that $\boldsymbol{\alpha}$ can be defined in infinitely many ways. Consider the following discounted pseudo-quadratic cost function
    \begin{align}\label{jjz}
        \mathcal{J}_z=\mathcal{J}+\int_{0}^{\infty} \exp({-2\gamma t})q_z(\boldsymbol{x},z)z^2\mathrm{d}t,
    \end{align}
where $\mathcal{J}$ is defined in \eqref{jjj}, and $q_z(\boldsymbol{x},z):\mathbb{R}^{n+1}\rightarrow\mathbb{R}^{>0}$ is a scalar $C^1$ function employed to assign a weight to the barrier state. Let us rewrite $\boldsymbol{y}$ in \eqref{output} as 
\begin{align}\label{sdcy}
    \boldsymbol{y}=\boldsymbol{H}(\boldsymbol{x})\boldsymbol{x},
\end{align}
where $\boldsymbol{H}:\mathbb{R}^{n}\rightarrow\mathbb{R}^{l\times n}$ is a non-unique function. Consider the following SDC representation for \eqref{eq20} and \eqref{eq20a}:
\begin{align}
\dot{\boldsymbol{v}}&=\boldsymbol{A}_\mathrm{d}(\boldsymbol{v})\boldsymbol{v},\label{asdc1}\\  \boldsymbol{y}_\mathrm{d}&=\boldsymbol{H}_\mathrm{d}(\boldsymbol{v})\boldsymbol{v},\label{asdcv1}
\end{align}
where \(\boldsymbol{A}_\mathrm{d}:\mathbb{R}^{n_\mathrm{d}}\rightarrow \mathbb{R}^{n_\mathrm{d} \times n_\mathrm{d}}\) and \(\boldsymbol{H}_\mathrm{d}:\mathbb{R}^{n_\mathrm{d}}\rightarrow \mathbb{R}^{l \times n_\mathrm{d}}\) are \(C^1\) functions. Define \(\boldsymbol{X} := e^{-\gamma t}[\boldsymbol{x}^\top, \,\boldsymbol{v}^\top,\, z]^\top\) and \(\boldsymbol{U} := e^{-\gamma t}\boldsymbol{u}\). Using \eqref{sdcf}, \eqref{sdcz}, and \eqref{asdc1}, the following SDC form for \(\boldsymbol{X}\) is derived:
\begin{align}\label{pp}
\dot{\boldsymbol{X}}=\boldsymbol{A}_z(\boldsymbol{X})\boldsymbol{X}+\boldsymbol{G}_z(\boldsymbol{X})\boldsymbol{U},
\end{align}
where 
\begin{align}\label{SDCfa}
        \boldsymbol{A}_{z}(\boldsymbol{X})&=\begin{bmatrix}
            \boldsymbol{A}(\boldsymbol{x})-\gamma \boldsymbol{I} & \boldsymbol{0} & 0\\
            \boldsymbol{0} & \boldsymbol{A}_\mathrm{d}(\boldsymbol{v})-\gamma \boldsymbol{I} & 0  \\
            \boldsymbol{\boldsymbol{\alpha}}^\top(\boldsymbol{x},z) & \boldsymbol{0} & -\gamma
        \end{bmatrix},\\
        \boldsymbol{G}_z(\boldsymbol{X})&=\begin{bmatrix}
            \boldsymbol{g}^\top(\boldsymbol{x}) & \boldsymbol{0} & \boldsymbol{\beta}(\boldsymbol{x},z)
        \end{bmatrix}^\top. \nonumber \\[-1cm]
\end{align}

We can also rewrite the cost function \eqref{jjz} as follows:
 \begin{align}\label{jjjt1}
        \mathcal{J}_z=\int_{0}^{\infty} \Big(\boldsymbol{X}^\top \boldsymbol{Q}_z(\boldsymbol{X})\boldsymbol{X}+\boldsymbol{U}^\top \boldsymbol{{R}}({\boldsymbol{x}})\boldsymbol{U}\Big)\mathrm{d}t,
    \end{align}
where 
\begin{equation}\label{Q_z}
\boldsymbol{Q}_z(\boldsymbol{X})=\mathrm{\textbf{diag}}\big(\boldsymbol{Q}_1(\boldsymbol{x},\boldsymbol{v}),q_z\big),    
\end{equation}
in which
\begin{eqnarray}\label{q1}
\boldsymbol{Q}_1\!(\boldsymbol{x},\boldsymbol{v})\!=\![\boldsymbol{H}(\boldsymbol{x}), -\boldsymbol{H}_\mathrm{d}(\boldsymbol{v})]^\top
\boldsymbol{Q}(\boldsymbol{x})[\boldsymbol{H}(\boldsymbol{x}), -\boldsymbol{H}_\mathrm{d}(\boldsymbol{v})]  \nonumber \\[-1cm]
\end{eqnarray}

The optimal solution to the infinite-horizon nonlinear optimal control problem defined by \eqref{pp} and \eqref{jjjt1} is given by $\boldsymbol{U}^\star=-\boldsymbol{R}^{-1}(\boldsymbol{x})\boldsymbol{G}_z^\top(\boldsymbol{X})\partial V^\top(\boldsymbol{X})/\partial\boldsymbol{X}$ where $V(\boldsymbol{X})$ represents the value function obtained by solving the associated Hamilton–Jacobi–Bellman (HJB) equation \cite{ccimen2008state}. However, solving the HJB equation analytically is generally too difficult or even impossible for most nonlinear systems except for relatively simple or low-dimensional cases. To address this challenge, the SDRE method is adopted as a practical approximation strategy. Through this approach, a suboptimal control law is obtained for the infinite-horizon optimal control problem posed by \eqref{pp} and \eqref{jjjt1}. It is important to emphasize that even if the original system described by \eqref{dynamics} and \eqref{output} is linear, the inclusion of the barrier state $z$ and its integration with the system dynamics produces a nonlinear augmented system (see Section \ref{sec4-2}). Using the SDRE technique, the following control law is taken:
   \begin{align}\label{SSDREt}
        \boldsymbol{U}(\!\boldsymbol{X})\!=\!-\!\boldsymbol{K}_z(\!\boldsymbol{X})\boldsymbol{X}\!=\!-\!\boldsymbol{R}^{-1}(\boldsymbol{x})\boldsymbol{G}_z^\top\!(\!\boldsymbol{X})\boldsymbol{P}_z(\!\boldsymbol{X})\boldsymbol{X},
    \end{align}
    where $\boldsymbol{P}_z:\mathbb{R}^{(n+n_\mathrm{d}+1)}\rightarrow\mathbb{R}^{(n+n_\mathrm{d}+1)\times (n+n_\mathrm{d}+1)}$ is the unique positive semi-definite solution of the SDRE
\begin{align}\label{sdrenewt}
&\boldsymbol{A}_z^\top(\boldsymbol{X})\boldsymbol{P}_z(\boldsymbol{X})+\boldsymbol{P}_z(\boldsymbol{X})\boldsymbol{A}_z(\boldsymbol{X})-\\\nonumber
&\boldsymbol{P}_z(\boldsymbol{X})\boldsymbol{G}_z(\boldsymbol{X})\boldsymbol{R}^{-1}(\boldsymbol{x})\boldsymbol{G}_z^\top(\boldsymbol{X})\boldsymbol{P}_z(\boldsymbol{X})=-\boldsymbol{Q}_z(\boldsymbol{X}).
\end{align}
The final control law is obtained as follows:
    \begin{align}\label{SSDREt1}
        \boldsymbol{u}=-\boldsymbol{K}_z(\boldsymbol{X})[\boldsymbol{x}^\top,\,\boldsymbol{v}^\top,\,z]^\top.
    \end{align}
From the control law \eqref{SSDREt1}, by rewriting \(\boldsymbol{K}_z(\boldsymbol{X}) = [\boldsymbol{K}_1(\boldsymbol{X}), \,\boldsymbol{K}_2(\boldsymbol{X}),\, K_3(\boldsymbol{X})]\), where \(\boldsymbol{K}_1(\boldsymbol{X}) \in \mathbb{R}^n\), \(\boldsymbol{K}_2(\boldsymbol{X}) \in \mathbb{R}^{n_\mathrm{d}}\), and \(K_3(\boldsymbol{X}) \in \mathbb{R}\), the proposed method yields a feedback-feedforward safe controller with gains \(\boldsymbol{K}_1\), \(\boldsymbol{K}_2\), and \(K_3\) computed simultaneously. The key assumption for deriving control law \eqref{SSDREt1} is the existence of a unique positive semi-definite solution to the SDRE \eqref{sdrenewt}. This holds if the triple \(\big(\boldsymbol{A}_z(\boldsymbol{X}), \boldsymbol{G}_z(\boldsymbol{X}), \boldsymbol{Q}_z^{1/2}(\boldsymbol{X})\big)\) is pointwise stabilizable and detectable \cite{ccimen2010systematic}. The following lemmas show that these conditions can be ensured by suitably choosing \(\gamma\).

\begin{assumption}\label{assumption_1}
   The pair \(\big(\boldsymbol{A}(\boldsymbol{x}), \boldsymbol{g}(\boldsymbol{x})\big)\) in \eqref{sdcf} is pointwise stabilizable for every \(\boldsymbol{x} \in \mathcal{S}\). 
\end{assumption}

\begin{lemma}
Under Assumption \ref{assumption_1}, \(\big(\boldsymbol{A}_z(\boldsymbol{X}), \boldsymbol{G}_z(\boldsymbol{X})\big)\) in \eqref{pp} is pointwise stabilizable provided that, for all \(\boldsymbol{v}\), the disscount factor \(\gamma\) satisfies
\begin{equation}\label{gamma_c}
\gamma > \max \Big( \Re \big(\lambda(\boldsymbol{A}_\mathrm{d}(\boldsymbol{v}))\big) \Big),
\end{equation}
where \(\Re\Big(\lambda\big(\boldsymbol{A}_\mathrm{d}(\boldsymbol{v})\big)\Big)\) shows the real parts of \(\lambda\big(\boldsymbol{A}_\mathrm{d}(\boldsymbol{v})\big)\).
\end{lemma}
\begin{pf}
Let us first construct the following matrix:
\begin{eqnarray*}
    \boldsymbol{\mathcal{M}}(\lambda)\!=\!\left[
    \setlength{\arraycolsep}{1.5pt} 
    \begin{array}{cccc}
        \!\!\boldsymbol{g}(\boldsymbol{x}) & \!\!(\lambda+\gamma)\boldsymbol{I}\!-\!\boldsymbol{A}(\boldsymbol{x}) & \!\!\boldsymbol{0} & \!\!0 \\
        \!\!\boldsymbol{0} & \!\!\boldsymbol{0} & \!\!(\lambda+\gamma) \boldsymbol{I}\!-\!\boldsymbol{A}_\mathrm{d}(\boldsymbol{v}) & \!\!0 \\ 
        \!\!\boldsymbol{\beta}^\top(\boldsymbol{X}) & \!\!-\boldsymbol{\alpha}^\top(\boldsymbol{X}) & \!\!\boldsymbol{0} & \!\!\lambda\!+\!\gamma
    \end{array}
    \right]
\end{eqnarray*}
The pair $\big(\boldsymbol{A}_z(\boldsymbol{X}),\boldsymbol{G}_z(\boldsymbol{X})\big)$ is pointwise stabilizable if $\boldsymbol{\mathcal{M}}(\lambda)$ is of full rank for all $\lambda\in\lambda\big({\boldsymbol{A}}_z(\boldsymbol{X})\big)$ with positive real part. Focusing on $\boldsymbol{A}_z(\boldsymbol{X})$ in \eqref{SDCfa}, we can conclude that $\lambda\big({\boldsymbol{A}}_z(\boldsymbol{X})\big)=\lambda\big(\boldsymbol{A}(\boldsymbol{X})-\gamma\boldsymbol{I}\big)\cup\lambda\big(\boldsymbol{A}_\mathrm{d}(\boldsymbol{v})-\gamma\boldsymbol{I}\big)\cup\{-\gamma\}$. Due to Assumption \ref{assumption_1}, $\boldsymbol{\mathcal{M}}(\lambda)$ is of full rank for $\lambda\in\lambda\big(\boldsymbol{A}(\boldsymbol{x})-\gamma\boldsymbol{I}\big)$ with positive real part. The second set comprises the eigenvalues of $\boldsymbol{A}_\mathrm{d}(\boldsymbol{v}) - \gamma \boldsymbol{I}$. Given that $\gamma$ satisfies the inequality \eqref{gamma_c} for all $\boldsymbol{v}$, the real part of $\lambda\in\lambda\big(\boldsymbol{A}_\mathrm{d}(\boldsymbol{v})-\gamma\boldsymbol{I}\big)$ is always negative. Lastly, since $\gamma > 0$, the final mode is similarly pointwise stabilizable. This completes the proof. 
\end{pf}

\begin{assumption}\label{assumption_2}
For \(\boldsymbol{A}(\boldsymbol{x})\), \(\boldsymbol{Q}(\boldsymbol{x})\), and \(\boldsymbol{H}(\boldsymbol{x})\) defined in \eqref{sdcf}, \eqref{jjz}, and \eqref{sdcy}, respectively, \(\big(\boldsymbol{A}(\boldsymbol{x}), \boldsymbol{Q}^{1/2}(\boldsymbol{x})\) \(\boldsymbol{H}(\boldsymbol{x})\big)\) is pointwise detectable for any \(\boldsymbol{x} \in \mathcal{S}\).
\end{assumption}
\begin{lemma}
Under Assumption 2, \(\big(\boldsymbol{A}_z(\boldsymbol{X}), \boldsymbol{Q}_z^{1/2}(\boldsymbol{X})\big)\), defined in \eqref{SDCfa} and \eqref{Q_z}, is pointwise detectable if for all \(\boldsymbol{v}\), \(\gamma\) satisfies the inequality \eqref{gamma_c}.
\end{lemma}
\begin{pf}
As concluded in the proof of Lemma 1, $\lambda\big({\boldsymbol{A}}_z(\boldsymbol{X})\big)\!=\!\lambda\big(\boldsymbol{A}(\boldsymbol{x})\!-\!\gamma\boldsymbol{I}\big)\cup\lambda\big(\boldsymbol{A}_\mathrm{d}(\boldsymbol{v})\!-\!\gamma\boldsymbol{I}\big)\cup\{-\gamma\}$.
By our assumption $\gamma$ is positive and $\gamma >  \max \Big( \Re \big( \lambda \left( \boldsymbol{A}_\mathrm{d} (\boldsymbol{v}) \right) \big) \Big)$ for all $\boldsymbol{v}$. Thus, \(\boldsymbol{A}_\mathrm{d}(\boldsymbol{v}) - \gamma \boldsymbol{I}\) is Hurwitz and \(-\gamma < 0\). Since $\big(\boldsymbol{A}(\boldsymbol{x}), \boldsymbol{Q}^{1/2}(\boldsymbol{x})\boldsymbol{H}(\boldsymbol{x})\big)$ is pointwise detectable for any $\boldsymbol{x}\in\mathcal{S}$, any unstable eigenvalue of $\boldsymbol{A}(\boldsymbol{x})$ is poitwise observable. As shifting the spectrum by \(\gamma\) does not alter the observability structure, any unstable eigenvalue of \(\boldsymbol{A}(\boldsymbol{x}) - \gamma I\) (i.e., with \(\Re(\lambda_i - \gamma) \geq 0\)) would correspond to an unstable eigenvalue of $\boldsymbol{A}(\boldsymbol{x})$ with \(\Re(\lambda_i) \geq \gamma\). Pointwise detectability of $\big(\boldsymbol{A}(\boldsymbol{x}), \boldsymbol{Q}^{1/2}(\boldsymbol{x})\boldsymbol{H}(\boldsymbol{x})\big)$ ensures that such modes are observable. The pointwise detectability of $\big(\boldsymbol{A}_\mathrm{d}(\boldsymbol{v}), \boldsymbol{Q}^{1/2}(\boldsymbol{x})\boldsymbol{H}_\mathrm{d}(\boldsymbol{v})\big)$ ensures that all unstable eigenvalues of $\boldsymbol{A}_\mathrm{d}(\boldsymbol{v})-\gamma I$ are either pointwise  observable or stable. However as $\boldsymbol{A}_\mathrm{d}(\boldsymbol{v}) - \gamma \boldsymbol{I}$ is Hurwitz, this requirement is trivially satisfied. Finally, the scalar eigenvalue \(-\gamma\) is stable and requires no pointwise observability for pointwise detectability. Thus, all unstable modes of $\boldsymbol{A}_z(\boldsymbol{X})$ are pointwise 
 observable via $\boldsymbol{Q}_z^{1/2}(\boldsymbol{X})$, establishing pointwise detectability of the pair $\big(\boldsymbol{A}_z(\boldsymbol{X}),\boldsymbol{Q}_z^{1/2}(\boldsymbol{X})\big)$ for all $\boldsymbol{x}\in\mathcal{S}$. 
\end{pf}
To assess the tracking performance and safe operation of the system \eqref{dynamics}–\eqref{output} under the proposed control law \eqref{SSDREt1}, we first state the following theorem.
\begin{theorem}\label{theorem2}
Suppose the discount factor \(\gamma\) is chosen for all \(\boldsymbol{v}\) such that inequality \eqref{gamma_c} holds. Further, assume that \(\boldsymbol{A}\), \(\boldsymbol{g}\), \(\boldsymbol{\alpha}\), \(\boldsymbol{\beta}\), \(\boldsymbol{Q}\), \(q_z\), and \(\boldsymbol{R}\) are all \(C^1\) functions. Then, given Assumptions \ref{assumption_1} and \ref{assumption_2}, the origin of system \eqref{pp} is asymptotically stable under the control law \eqref{SSDREt1}.
\end{theorem}
\begin{pf}
Based on Lemmas 1 and 2, we can deduce that the triple \(\big(\boldsymbol{A}_z(\boldsymbol{X}),\boldsymbol{G}_z(\boldsymbol{X}), \boldsymbol{Q}_z^{1/2}(\boldsymbol{X})\big)\) is pointwise stabilizable and detectable for any $\boldsymbol{x}\in\mathcal{S}$. Therefore, the SDRE \eqref{sdrenewt} has a unique positive semi-definite solution $\boldsymbol{P}_z(\boldsymbol{X})$. Using Theorem 1, we can demonstrate that the origin of the augmented system \eqref{pp} under the control law \eqref{SSDREt1} is asymptotically stable. 
\end{pf}
\begin{remark}\label{remark_safe}
According to Theorem 2, under the proposed SSDRE tracking controller, \(\boldsymbol{X}\) converges asymptotically to zero. As a result, the tracking error asymptotically vanishes when the discount factor \(\gamma\) is sufficiently small. Meanwhile, \(z\) remains bounded for small \(\gamma\), ensuring the safety constraints in \eqref{si} are satisfied. Thus, an open set \(\mathcal{S}_z \subseteq \mathcal{S}\) exists, which is forward invariant. Notably, \(\gamma\) can be selected small enough for common reference trajectories, including sinusoidal and step-like signals \textup{(}see Sections \ref{sec4-2} and \ref{sec:experiment}\textup{)}.
\end{remark}

\begin{remark}
When implementing the proposed SSDRE method, we can also construct an SDC representation for the system described by \eqref{dynamics}-\eqref{output} and \eqref{eq20}-\eqref{eq20a}. It is important to note that the conclusions of Theorem 2 and Remark \ref{remark_safe} remain valid, provided the chosen SDC representation for the system \eqref{dynamics}-\eqref{output} and \eqref{eq20}-\eqref{eq20a} is pointwise stabilizable and detectable for \(\boldsymbol{x} \in \mathcal{S}\) \textup{(}see Section \ref{sec:experiment}\textup{)}.
\end{remark}

\begin{remark}\label{remark_qz}
In our control design, safety constraints take precedence over tracking performance. In case of conflict, the controller prioritizes safety, potentially at the cost of tracking accuracy \textup{(}see Sections \ref{conf} and \ref{sec4-2}\textup{)}. This behavior is guaranteed when \(q_z \in \mathbb{R}^{>0}\). Setting \(q_z = 0\) may decouple the barrier state \(z\) from the control input. For instance, if \(\boldsymbol{\beta}(\boldsymbol{x}, z)\) in \eqref{sdcz} is zero, as in Sections \ref{sec4-1} and \ref{sec:experiment}, the control law derived from \eqref{SSDREt1} becomes independent of \(z\), i.e., $K_3=0$, and safety can no longer be ensured. Only when \(\boldsymbol{\alpha}(\boldsymbol{x}, z) \neq \boldsymbol{0}\) or \(\boldsymbol{\beta}(\boldsymbol{x}, z) \neq \boldsymbol{0}\) with \(q_z \in \mathbb{R}^{>0}\) does the barrier state actively influence the control input, ensuring the system remains within the safe set.
\end{remark}

\begin{remark}\label{remark_new_z}
   When the safety requirements and tracking objectives are not conflicting \textup{(}see Section~\ref{sec:experiment}\textup{)}, $p$ in \eqref{smct_1} can be defined as a function of the tracking error $\boldsymbol{e}$. In this case, the SDC given in \eqref{sdcz} takes the form
\begin{equation}\label{sdcxdd}
\dot{z} \!=\! \boldsymbol{\alpha}^\top\!(\boldsymbol{x}, \boldsymbol{x}_\mathrm{d}, z) \boldsymbol{x}\! +\! \boldsymbol{\alpha}_\mathrm{d}^\top\!(\boldsymbol{x}, \boldsymbol{x}_\mathrm{d}, z) \boldsymbol{x}_\mathrm{d} \!+\! \boldsymbol{\beta}^\top\!(\boldsymbol{x}, \boldsymbol{x}_\mathrm{d}, z) \boldsymbol{u},
\end{equation}
where $\boldsymbol{\alpha}$, $\boldsymbol{\alpha}_\mathrm{d}$, and $\boldsymbol{\beta}$ are appropriately dimensioned vectors. It can be shown that, by employing \eqref{sdcxdd} in the SDC form \eqref{SDCfa}, the results established in Lemmas~1 and 2 as well as Theorem~2 remain valid. 
\end{remark}

\subsection{Solution with $N$ Barrier States}\label{multis}
In this section, we extend the proposed SSDRE controller to address the safe tracking problem by introducing \(N\) individual barrier states corresponding to the safety constraints in \eqref{si}. First, we define 
\begin{align}
 z_i(t):=\frac{p_i\big(\boldsymbol{x}(t)\big)}{q_i\Big(s_i\big(\boldsymbol{x}(t)\big)\Big)}   
\end{align}
for \(i=1, \dots, N\), where \(p_i:\mathbb{R}^n \rightarrow \mathbb{R}\) and \(q_i:\mathbb{R} \rightarrow \mathbb{R}\) are smooth functions for \(\boldsymbol{x} \in \mathcal{S}_i\), satisfying \(p_i(\boldsymbol{0})=q_i(0)=0\). Using  \eqref{sdcf}, the dynamics of \(z_i\) can be expressed as
\begin{align}\label{ggk1}
    \dot{z}_i\!=\!\Big(\!\frac{\partial p_i(\boldsymbol{x})}{\partial \boldsymbol{x}}\!-\!z_i\frac{\partial q_i\big(s_i(\boldsymbol{x})\big)}{\partial s_i(\boldsymbol{x})}\frac{\partial s_i(\boldsymbol{x})}{\partial \boldsymbol{x}}\!\Big)^\top\!\frac{\dot{\boldsymbol{x}}}{q_i\big(s_i(\boldsymbol{x})\big)}.
\end{align}
Let the right-hand side of \eqref{ggk1} be expressed in its SDC form as follows:
\begin{align}\label{zsd}
\dot{z}_i=\boldsymbol{\alpha}_i^\top(\boldsymbol{x},z_i)\boldsymbol{x}+\boldsymbol{\beta}_i^\top(\boldsymbol{x},z_i)\boldsymbol{u},   
\end{align}
where $\boldsymbol{\alpha}_i$ and $\boldsymbol{\beta}_i$ have proper dimensions. 
By defining $\boldsymbol{X}:=\exp{(-\gamma t)}[\boldsymbol{x}^\top,\,\boldsymbol{v}^\top,\,z_1,\,\dots,\,z_N]^\top$ and $\boldsymbol{U}:=\exp{(-\gamma t)}\boldsymbol{u}$, and using \eqref{asdc1} and \eqref{zsd}, an SDC representation of the form \eqref{pp} is obtained with
\begin{align}
\boldsymbol{A}_z(\boldsymbol{X})&=\begin{bmatrix}
        \!\boldsymbol{A}(\boldsymbol{x})\!-\!\gamma I & \!\boldsymbol{0} & \!0&\!\dots &\!0 & \!0\\
     \!\boldsymbol{0} & \!\boldsymbol{A}_\mathrm{d}(\boldsymbol{v})\!-\!\gamma I & \!0& \!\dots &\! 0 & \! 0\\
     \!\boldsymbol{\alpha}^\top_1(\boldsymbol{x},z_1)&\!\boldsymbol{0} & \!-\gamma & \!\dots& \! 0 & \! 0\\
     \!\vdots & \!\vdots  &\! \vdots &\! \vdots& \!\vdots&\!\vdots\\
     \!\boldsymbol{\alpha}^\top_N(\boldsymbol{x},z_N)&\!\boldsymbol{0} & \!0&\! \dots&\! 0& \!-\gamma
       \end{bmatrix},\label{ghf}\\
\boldsymbol{G}_z(\boldsymbol{X})&=\begin{bmatrix}
               \boldsymbol{g}\!^\top(\boldsymbol{x}) &
               \boldsymbol{0}&               \boldsymbol{\beta}_1^\top(\boldsymbol{x},z_1)  & \dots & \boldsymbol{\beta}_N^\top(\boldsymbol{x},z_N)
           \end{bmatrix}^\top.\label{ghg}
\end{align}
The corresponding cost function for the augmented system with $N$ barrier states is defined as:
    \begin{align}\label{jjzn}
        \mathcal{J}_z=\mathcal{J}+\int_{0}^{\infty} \exp({-2\gamma t})\Big(\sum_{i=1}^{N}q_{z_i}(\boldsymbol{x},z_i)z_i^2\Big)\mathrm{d}t,
    \end{align}
where \(q_{z_i}(\boldsymbol{x}, z_i): \mathbb{R}^{n+1} \rightarrow \mathbb{R}^{>0}\) for \(i=1, \dots, N\) is a \(C^1\) function used to assign a weight to the barrier state \(z_i\). We can rewrite \(\mathcal{J}_z\) in \eqref{jjzn} as
 \begin{align}\label{jjjt2}
        \mathcal{J}_z=\int_{0}^{\infty} \Big(\boldsymbol{X}^\top \boldsymbol{Q}_z(\boldsymbol{X})\boldsymbol{X}+\boldsymbol{U}^\top \boldsymbol{{R}}({\boldsymbol{x}})\boldsymbol{U}\Big)\mathrm{d}t,
    \end{align}
where 
\begin{align}\label{qzn}
\boldsymbol{Q}_z(\!\boldsymbol{X})\!=\!\mathrm{\textbf{diag}}\big(\boldsymbol{Q}_1\!(\boldsymbol{x},\!\boldsymbol{v}),q_{z_1}\!(\boldsymbol{x},\!z_1),\dots,q_{z_N}\!(\boldsymbol{x},\!z_N)\big)
\end{align}
in which $\boldsymbol{Q}_1(\boldsymbol{x},\boldsymbol{v})$ is defined by \eqref{q1}. By employing the SDRE control technique, the control law \(\boldsymbol{u} = -\boldsymbol{R}^{-1}(\boldsymbol{x}) \boldsymbol{G}_z^\top(\boldsymbol{X}) \boldsymbol{P}_z(\boldsymbol{X}) \boldsymbol{X}\) is derived, where \(\boldsymbol{P}_z(\boldsymbol{X}) \in \mathbb{R}^{(n+n_\mathrm{d}+N)\times (n+n_\mathrm{d}+N)}\) is the unique positive semi-definite solution of the SDRE \eqref{sdrenewt}, with \(\boldsymbol{A}_z(\boldsymbol{X})\) and \(\boldsymbol{G}_z(\boldsymbol{X})\) defined in \eqref{ghf} and \eqref{ghg}, respectively.

\begin{corollary}\label{corollary_2}
Under Assumptions \ref{assumption_1} and \ref{assumption_2}, the triple $\big(\boldsymbol{A}_z(\boldsymbol{X}), \boldsymbol{G}_z(\boldsymbol{X}), \boldsymbol{Q}_z^{1/2}(\boldsymbol{X})\big)$, defined in \eqref{ghf}, \eqref{ghg}, and \eqref{qzn}, is pointwise stabilizable and detectable, provided that for all $\boldsymbol{v}$ the discount factor $\gamma$ satisfies \eqref{gamma_c}.
\end{corollary}
\begin{pf}
The proof can be obtained by extending the approach used in the proofs of Lemmas 1 and 2.
\end{pf}

We can also investigate the satisfaction of the control objectives by the following corollary. 
\begin{corollary}
Suppose Assumptions \ref{assumption_1} and \ref{assumption_2} hold, and that for all $\boldsymbol{v}$ the discount factor $\gamma$ satisfies inequality \eqref{gamma_c}. Then the origin of system \eqref{pp} with $\boldsymbol{A}_z(\boldsymbol{X})$ and $\boldsymbol{G}_z(\boldsymbol{X})$ (defined in \eqref{ghf} and \eqref{ghg}, respectively) is asymptotically stable under the control law $        \boldsymbol{u}=-\boldsymbol{K}_z(\boldsymbol{X})[\boldsymbol{x}^\top,\,\boldsymbol{v}^\top,\,z_1,\,\dots,\,z_N]^\top$. 
    \end{corollary}
\begin{pf}
The steps to prove this theorem are identical to those used in the proof of Theorem 2.
\end{pf}

\begin{remark}
Consistent with the conclusion drawn in Remark \ref{remark_safe}, it can be asserted that for the proposed solution incorporating \(N\) barrier states, if \(\gamma\) is selected sufficiently small, there exists an open set \(\mathcal{S}_z \subseteq \mathcal{S}\) that remains forward invariant, thereby ensuring the satisfaction of the safety conditions specified in \eqref{si}. However, as indicated by Lemmas 1--3, the choice of \(\gamma\) must also account for the dynamics of the desired trajectory. For example, in the scenario considered in Section \ref{non_1}, it is required that \(\gamma > 0.5\). The proposed SSDRE controller offers notable flexibility in tuning through the weighting matrices \(\boldsymbol{Q}(\boldsymbol{x})\) and \(\boldsymbol{R}(\boldsymbol{x})\). Specifically, increasing \(\boldsymbol{Q}(\boldsymbol{x})\) enhances the convergence rate of the tracking error. 
\end{remark}

%% file: simulation.tex
\section{Simulation Results}\label{simulation}
In this section, the effectiveness of the proposed SSDRE controller is evaluated through two case studies. All simulations were conducted using \textsc{Matlab} 2023a on a desktop computer equipped with an Intel\textregistered{} Core\texttrademark{} i5-10400F CPU running at 2.9~GHz and 8~GB of RAM. 
\subsection{Case Study 1: A Mechanical System}\label{sec4-1}
Consider the following nonlinear system \cite{dong2024safety}:
\begin{align}\label{case2}
\dot{x}_1&=x_3,\nonumber\\
\dot{x}_2&=x_4,\\
\dot{x}_3&=\!-\!(0.8+0.2\exp{(-100\vert x_3\vert})\mathrm{tanh}(x_3)\!-\!x_3\!-\!x_1+u_1,\nonumber\\
\dot{x}_4&=\!-\!(0.8+0.2\exp{(-100\vert x_4\vert})\mathrm{tanh}(x_4)\!-\!x_4\!-\!x_2+u_2.\nonumber
\end{align}
In the rest of this Section \ref{sec4-1}, the SDC representation $\dot{\boldsymbol{x}}=\boldsymbol{A}(\boldsymbol{x})\boldsymbol{x}+\boldsymbol{g}\boldsymbol{u}$ is considered with $\boldsymbol{g}=[\boldsymbol{0},\,\boldsymbol{I}_2]^\top$ and
\begin{align}\label{case2_A}
\boldsymbol{A}(\boldsymbol{x})&=\begin{bmatrix}
    0 & 0 & 1 & 0\\
    0 & 0 & 0 & 1\\
    -1 & 0 & a_{33}(x_3) & 0\\
    0 & -1 & 0 & a_{44}(x_4)
\end{bmatrix},
\end{align}
where 
\begin{align*}
\begin{split}
a_{33}(x_3)&=-(0.8+0.2\exp{(-100\vert x_3\vert})\frac{\mathrm{tanh}(x_3)}{x_3}-1,\\
a_{44}(x_4)&=-(0.8+0.2\exp{(-100\vert x_4\vert})\frac{\mathrm{tanh}(x_4)}{x_4}-1.
\end{split}
\end{align*}
It should be noted that since \(\lim_{x \to 0} \tanh(x)/x = 1\), the functions \(a_{33}(x_3)\) and \(a_{44}(x_4)\) are well-defined. By constructing the state-dependent controllability matrix \(\boldsymbol{\Phi}_c = [\boldsymbol{g},\, \boldsymbol{A}(\boldsymbol{x})\boldsymbol{g},\, \boldsymbol{A}^2(\boldsymbol{x})\boldsymbol{g},\, \boldsymbol{A}^3(\boldsymbol{x})\boldsymbol{g}]\), it can be verified that the pair \(\big(\boldsymbol{A}(\boldsymbol{x}), \boldsymbol{g}\big)\) is pointwise controllable. The system output is defined as \(\boldsymbol{y} \coloneqq [x_1,\, x_2]^\top\). Accordingly, by considering \(\boldsymbol{H} = [\boldsymbol{I}_2,\, \boldsymbol{0}]\) as per \eqref{sdcy}, the pair \(\big(\boldsymbol{A}(\boldsymbol{x}), \boldsymbol{Q}^{1/2}(\boldsymbol{x})\boldsymbol{H}\big)\) is pointwise observable, provided that \(\boldsymbol{Q}(\boldsymbol{x})\) is positive definite. 

In this case study, three different tracking problems are addressed. The first, referred to as the non-conflicted case, involves applying the proposed method when no conflict arises between the safety and tracking objectives. The second, termed the conflicted case, considers scenarios where the safety requirements oppose the tracking objectives, and our approach is employed to manage this conflict. Finally, a scenario involving three safety constraints is investigated.

\subsubsection{Non-conflicted case}\label{non_1}
In this case, the system output is required to track the desired trajectory \(\boldsymbol{y}_\mathrm{d} \coloneqq \boldsymbol{v} = [v_1,\, v_2]^\top\), which is governed by the following dynamics \cite{dong2024safety}:
\begin{align}\label{case2_V1}
\begin{split}
\dot{v}_1&=v_2,\\
\dot{v}_2&=-v_1+(1-v_1^2)v_2.
\end{split}
\end{align}
As per \eqref{asdc1} and \eqref{asdcv1}, the system described by \eqref{case2_V1} can be represented in the SDC form as follows:
\begin{align}\label{case2_V1_sdc}
\begin{split}
\dot{\boldsymbol{v}}&=\begin{bmatrix}
    0 & 1\\
    -1 & 1-v_1^2
\end{bmatrix}\boldsymbol{v}=\boldsymbol{A}_\mathrm{d}(v_1)\boldsymbol{v},\\
\boldsymbol{y}_\mathrm{d} &=\boldsymbol{I}_2\boldsymbol{v}=\boldsymbol{H}_\mathrm{d}\boldsymbol{v}.
\end{split}
\end{align}
Consider the initial condition $\boldsymbol{v}_0 = [2,\,2]^\top$. With this initial condition, $v_1(t)$ remains within the interval $(-2.02,\,2.32)$ for all $t \geq 0$ (see Fig.~\ref{Fig_1}). The eigenvalues of $\boldsymbol{A}_\mathrm{d}(v_1)$ satisfy
\[
\max \Big( \Re \big( \lambda ( \boldsymbol{A}_\mathrm{d} (v_1) ) \big) \Big) \leq 0.5,
\]
and hence, the discount factor must satisfy $\gamma > 0.5$ in accordance with Lemmas~1--3.
Assume the system must satisfy the safety constraints $\vert x_i\vert<3$ for $i=1,2$ \cite{dong2024safety}. Thus, the safe set is defined as $\mathcal{S}=\mathcal{S}_1\cap\mathcal{S}_2$, where $\mathcal{S}_i\coloneqq\{\boldsymbol{x}\in\mathbb{R}^4 | s_i=9-x_i^2>0\}$ for $i=1,2$. We address this problem by applying the method proposed in Section \ref{singles}. To this end, a single barrier state is defined as $z\coloneqq (x_1^2+x_2^2)/s$, where $s=s_1s_2$. Considering the system dynamics given in \eqref{case2}, we have
\begin{equation}\label{case1_barrier_dynamics}
    \dot{z}=\frac{2x_1x_3+2x_2x_4}{s_1s_2}+\frac{2zx_1x_3}{s_1}+\frac{2zx_2x_4}{s_2}
\end{equation}
As per \eqref{sdcz}, we have $\boldsymbol{\beta}=\boldsymbol{0}$, and we select $\boldsymbol{\alpha}=[2zx_3/s_1,\ 2zx_4/s_2,\ 2x_1/s,\ 2x_2/s]^\top$. We also set $\gamma=0.6$, $\boldsymbol{Q}=10^3/(\Vert \boldsymbol{e}\Vert^2+10^{-3})\boldsymbol{I}_2$, $\boldsymbol{R}=\boldsymbol{I}_2$, $q_z=1$ as per equation \eqref{jjz}. In Fig.~\ref{Fig_1}(a), for \(\boldsymbol{x}_0 = [2.5,\ -2.5,\ 5,\ 2]^\top\), the results of applying the proposed method (blue solid line) are compared with those from a conventional SDRE controller using the same state and input weighting matrices \(\boldsymbol{Q}\) and \(\boldsymbol{R}\) (dotted black line). Additionally, results obtained using a nonlinear MPC are included for comparison. In the nonlinear MPC setup \cite{Grune2017}, the system dynamics given by \eqref{case2} and \eqref{case2_V1} are discretized with a sampling time of $0.01$ seconds, and the prediction horizon is set to $6$ steps. It is observed that while the proposed method successfully enforces the safety constraints, the conventional SDRE controller, configured with \(q_z = 0\) (as discussed in Remark \ref{remark_qz}), fails to do so. Since the safety constraints and tracking objectives are not in conflict for this scenario, the SSDRE controller achieves effective trajectory tracking, with the reference trajectory illustrated by the dashed red line in Fig.~\ref{Fig_1}(a). The system trajectory resulting from the nonlinear MPC (represented by the orange dash-dotted line) also satisfies the safety constraints. The corresponding control inputs generated by the proposed method are shown in Fig.~\ref{Fig_1}(b), where a zoomed-in section highlights the smoothness of the computed control actions. Based on the simulation, the system trajectory enters the region \(\Vert \boldsymbol{e} \Vert < 0.1\) within \(0.69\) seconds, while the SDRE and nonlinear MPC require \(0.8\) and \(1.22\) seconds, respectively. By defining the performance index \( J_{\boldsymbol{e}} \coloneqq \int_0^{10} \Vert \boldsymbol{e}(t) \Vert \, \mathrm{d}t \), the resulting values of this index are \(1.34\), \(1.48\), and \(2.63\) for the SSDRE, SDRE, and nonlinear MPC, respectively. Furthermore, the SDRE controller is solved at a sampling rate of $100$ Hz. The proposed controller, which involves solving the Riccati equation \eqref{sdrenewt} at this rate, required approximately $0.8$ seconds to simulate $10$ seconds of system operation. In comparison, the simulation time for the nonlinear MPC was approximately $3.7$ seconds. Lastly, based on our simulations, it was observed that the tracking objective could not be reliably achieved when the prediction horizon was set to fewer than $6$ steps. 

\input{figures/simulation_01}

\input{figures/simulation_02}
\subsubsection{Conflicted case}\label{conf}
Assume the desired trajectory is generated by 
\begin{align}\label{conflict}
\begin{split}
\dot{\boldsymbol{v}}=\begin{bmatrix}
    0 & 1 & 0 & 0\\
    -1 & -1 & 0 & 1\\
    0 & 0 & 0 & 1\\
    0 & 0 & -1 & 0\\
\end{bmatrix}\boldsymbol{v}=\boldsymbol{A}_\mathrm{d}\boldsymbol{v}.
\end{split}
\end{align}
With $\boldsymbol{y}_\mathrm{d} = [v_1,\, v_2]^\top$, we define $\boldsymbol{H}_\mathrm{d} = [\boldsymbol{I}_2,\, \boldsymbol{0}]$. The safety constraints are identical to those in the non-conflicted case, namely $\vert x_i(t)\vert < 3$ for $i=1,2$ and all $t \geq 0$. To address this problem, the proposed SSDRE method is applied, utilizing a single barrier state from Section \ref{singles} and two barrier states from Section \ref{multis}. In both cases, since $\lambda\big(\boldsymbol{A}_\mathrm{d}\big)=\{-0.5 \pm i\sqrt{3}/2,\ \pm i\}$, the discount factor $\gamma$ can be selected as any positive constant satisfying the conditions of the proposed framework and Lemmas~1–3. We choose $\gamma = 0.01$ and set the weighting matrices as $\boldsymbol{Q} = 10^3 / (\Vert \boldsymbol{e} \Vert^2 + 10^{-3}) \boldsymbol{I}_2$ and $\boldsymbol{R} = \boldsymbol{I}_2$.

In the SSDRE control design with a single barrier state, the variable $z$ and its SDC form are defined as in Section~\ref{non_1}. For $\boldsymbol{x}_0 = [1,\ 1,\ -6,\ -5]^\top$ and $\boldsymbol{v}_0 = [2,\ 2,\ -2,\ 3]^\top$, the simulation results using the proposed SSDRE approach are shown in Fig.~\ref{Fig_2} for two values of \(q_z\): \(10^2\) and \(10^4\). As illustrated, the safety constraints are satisfied in both cases. However, increasing $q_z$ makes the closed-loop system more conservative, as seen by comparing the blue solid line (\(q_z=10^2\)) and the orange dash-dotted line (\(q_z=10^4\)), with the latter causing greater deviation from the desired trajectory. In contrast, the conventional SDRE controller (black dotted line) maintains trajectory tracking but violates safety constraints. This result confirms that the  SSDRE controller prioritizes safety over trajectory tracking.

We also address this problem using an SSDRE controller designed following the method in Section~\ref{multis}. Accordingly, two barrier states are defined as $z_i \coloneqq (x_1^2 + x_2^2)/s_i$, where $s_i = 9 - x_i^2$ for $i=1,2$. The corresponding dynamics of these barrier states are computed as follows:
\begin{equation}\label{barrier_dynamics_2}
    \dot{z}_i=\frac{2x_1x_3+2x_2x_4+2z_ix_ix_{i+2}}{s_i}.
\end{equation}
In the SDC form~\eqref{zsd}, we set $\boldsymbol{\beta}_1 = \boldsymbol{\beta}_2 = \boldsymbol{0}$ and choose $\boldsymbol{\alpha}_1 = [0,\, 0,\, 2x_1(1+z_1)/s_1,\ 2x_2/s_1]$ and $\boldsymbol{\alpha}_2 = [0,\, 0,\, 2x_1/s_2,\, 2x_2(1+z_2)/s_2]$. Fig.~\ref{Fig_3} shows the system trajectories for three different combinations of $q_{z_1}$ and $q_{z_2}$. Although the safety constraints are satisfied in all cases, the results highlight how the system’s conservativeness relative to the safety constraints varies. For example, comparing the blue solid line ($q_{z_1}=q_{z_2}=10$) with the orange dash-dotted line ($q_{z_1}=100$, $q_{z_2}=10$) reveals that increasing $q_{z_1}$, associated with the constraint $\vert x_1 \vert < 3$, makes the system more conservative with respect to this limit, without notably affecting $\vert x_2 \vert < 3$. A similar trend is observed when increasing $q_{z_2}$ by comparing the blue solid line with the black dotted line. These results confirm that, by introducing multiple barrier states and applying the method in Section~\ref{multis}, the conservativeness of the system’s response can be tuned independently for each safety constraint — offering  flexibility in controller design.
\input{figures/simulation_03}

\subsubsection{Tracking problem with three safety constraints}\label{case_1_three}
Consider the desired trajectory generated by the system \eqref{case2_V1}. The objective is to design a tracking controller that ensures the forward invariance of the following sets:
\begin{align*}
\begin{split}
\mathcal{S}_j&=\left\{ \boldsymbol{x} \in \mathbb{R}^4 \ \middle| \ s_j(\boldsymbol{x})\coloneqq 9-x_j^2> 0 \right\},\quad\!\!\!\!\textup{for}\quad\!\!\!\!j=1,2\\
\mathcal{S}_3&=\left\{ \boldsymbol{x} \in \mathbb{R}^4 \ \middle| \ s_3(\boldsymbol{x})\coloneqq x_1^2+x_2^2> 0.5^2 \right\}.
\end{split}
\end{align*}
The first two constraints limit the amplitudes of \(x_1\) and \(x_2\), while the condition \(s_3 > 0\) prevents the system trajectory from entering a circular region centered at the origin with a radius of $0.5$. To address this problem, an SSDRE controller is designed using a single barrier state defined as $z\coloneqq5(x_1^2+x_2^2)/\big(s_1(\boldsymbol{x})s_2(\boldsymbol{x})s_3(\boldsymbol{x})\big)$. 
The dynamics of this barrier state are derived as follows:
\begin{align*}
    \frac{\dot{z}}{10}=\frac{x_1x_3\!+\!x_2x_4}{s_1s_2s_3}+\frac{zx_1x_3}{s_1}+\frac{zx_2x_4}{s_2}
    -\frac{z(x_1x_3\!+\!x_2x_4)}{s_3}.
\end{align*}
We set \(\boldsymbol{\beta} = \boldsymbol{0}\) and choose
$\boldsymbol{\alpha}=[10zx_3/s,\ 10zx_4/s,$ $10zx_1(1/s_1-1/s_3),\ 10zx_2(1/s_2-1/s_3)]$.
Following the non-conflicted case in Section~\ref{non_1}, we set \(\gamma = 0.6\), \(\boldsymbol{Q} = 10^3/(\Vert \boldsymbol{e} \Vert^2 + 10^{-3}) \boldsymbol{I}_2\), \(\boldsymbol{R} = \boldsymbol{I}_2\), and \(q_z = 100\) in \eqref{jjz}. For \(\boldsymbol{v}_0 = [2,\, 2]^\top\) with \(\boldsymbol{x}_1(0) = [2.5,\, -2.5,\, 5,\, 2]^\top\) and \(\boldsymbol{x}_2(0) = [-1,\, 1,\, 5,\, -5]^\top\), the SSDRE controller’s performance is depicted in Fig.~\ref{Fig_4}. As illustrated, the proposed controller effectively enforces the safety constraints. Moreover, the tracking objective is achieved, as the tracking error \(\boldsymbol{e}\) reaches and remains within the bound \(\Vert \boldsymbol{e} \Vert < 0.1\) in \(1.03\) seconds and \(1.14\) seconds for the initial conditions \(\boldsymbol{x}_1(0)\) and \(\boldsymbol{x}_2(0)\), respectively.

\input{figures/simulation_04}

\subsection{Case Study 2: Mobile Robots Collision Avoidance}\label{sec4-2}
In this case study, two simple mobile robots are required to follow predefined time-varying desired trajectories. The SSDRE tracking controller is applied to ensure that, in addition to accurately tracking these trajectories, (i) each robot avoids collisions with obstacles, and (ii) collisions between the robots are also prevented. The dynamics of the robots for \(j=1,2\) are given by \cite{romdlony2016stabilization,almubarak2023barrier}:
\begin{align}\label{case1}
\begin{split}
\dot{x}_{j,1}(t)&=u_{j,1}(t),\\
\dot{x}_{j,2}(t)&=u_{j,2}(t),
\end{split}
\end{align}
where $x_{j,1}$ and $x_{j,2}$ denote the positions of the $j^{\mathrm{th}}$ robot in a 2D plane, while $u_{j,1}$ and $u_{j,2}$ represent its velocities. The tracking and safety objectives are defined as follows:

$i)$ \textbf{Trajectory tracking:} Each robot is required to follow a circular path centered at the origin with a radius of \(2\,\mathrm{m}\). The first robot completes the loop every \(2\pi\) seconds, while the second completes it every \(4\pi\) seconds.

$ii)$ \textbf{Obstacle avoidance:} The robots must avoid a circular obstacle centered at \((2,2)\,\mathrm{m}\) with radius \(1.5\,\mathrm{m}\).

$iii)$ \textbf{Collision avoidance:} The robots must maintain a minimum separation of \(\delta = 0.1\,\mathrm{m}\) to avoid collisions.

Note that a conflict between the tracking objective and obstacle avoidance is evident in Fig.~\ref{Fig_5}(a). Define \(\boldsymbol{x}_j \coloneqq [x_{j,1},\, x_{j,2}]^\top\) for \(j=1,2\), and \(\boldsymbol{x} \coloneqq [\boldsymbol{x}_1^\top,\, \boldsymbol{x}_2^\top]^\top\). In the remainder of this section, we consider $\boldsymbol{A}=\boldsymbol{0}$ and $\boldsymbol{g}=\boldsymbol{I}_4$ according to \eqref{sdcf}, and set $\boldsymbol{H}=\boldsymbol{I}_4$ as specified in \eqref{sdcy}. Under these conditions, $(\boldsymbol{A},\boldsymbol{g},\boldsymbol{H})$ is both controllable and observable. The desired trajectories are given by
\[
v_{j,1} = 2 \sin(\omega_j t), \quad v_{j,2} = 2 \cos(\omega_j t), \quad j=1,2,
\]
with \(\omega_1 = 1\,\mathrm{rad/sec}\) and \(\omega_2 = 0.5\,\mathrm{rad/sec}\), where \(v_{j,1}\) and \(v_{j,2}\) correspond to the desired positions for \(x_{j,1}\) and \(x_{j,2}\), respectively. Setting \(\boldsymbol{y}_\mathrm{d} = \boldsymbol{v} \coloneqq [v_{1,1},\, v_{1,2},\, v_{2,1},\, v_{2,2}]^\top\) with initial condition \(\boldsymbol{v}_0 = [0,\, 2,\, 0,\, 2]^\top\), the following dynamical system generates the desired trajectories:
\begin{align}\label{mobile_ad}
\begin{split}
\dot{\boldsymbol{v}}=\begin{bmatrix}
    0 & 1 & 0 & 0\\
    0 & -1 & 0 & 0\\
    0 & 0 & 0 & 1\\
    0 & 0 & -0.25 & 0\\
\end{bmatrix}\boldsymbol{v}=\boldsymbol{A}_\mathrm{d}\boldsymbol{v}. \\[-0.5cm]
\end{split}
\end{align}

As specified in \eqref{asdcv1}, we set $\boldsymbol{H}_\mathrm{d} = \boldsymbol{I}_4$. To formulate the safety objectives, $\mathcal{S}_j$ for $j=1,2$ and $\mathcal{S}_3$ are defined as
\\[-1cm]
\begin{align*}
\begin{split}
\mathcal{S}_j\!&=\!\left\{ \boldsymbol{x} \!\in \!\mathbb{R}^4 | \!\ s_j\!\coloneqq \!(x_{j,1}\!-\!2)\!^2\!+\!(x_{j,2}\!-\!2)\!^2\!-\!1.5\!^2\!>\! 0 \!\right\}, \\
\mathcal{S}_3&\!=\!\left\{ \boldsymbol{x} \!\in\! \mathbb{R}^4 | \!\ s_3\!\coloneqq \!(x_{1,1}\!-\!x_{2,1})\!^2\!+\!(x_{1,2}\!-\!x_{2,2})\!^2\!-\!0.1\!^2>\! 0 \!\right\}. \\[-1cm]
\end{split}
\end{align*}

We address this problem using the SSDRE controller proposed in Section~\ref{multis}. By defining $z_j\coloneqq(x_{j,1}^2+x_{j,2}^2)/s_j(\boldsymbol{x}_j)$ for $j=1,2$ and $z_3\coloneqq\Vert \boldsymbol{x}_1-\boldsymbol{x}_2\Vert^2/s_3(\boldsymbol{x})$, we set  $\boldsymbol{\alpha}_i = \boldsymbol{0}$ for $i=1,2,3$, and we have $\boldsymbol{\beta}_1=[\beta_{1,1},\,\beta_{1,2},\,0,\,0]$, $\boldsymbol{\beta}_2=[0,\,0,\,\beta_{2,1},\,\beta_{2,2}]$, and $\boldsymbol{\beta}_3=[\beta_{3,1},\,\beta_{3,2},\,\beta_{3,3},\,\beta_{3,4}]$ as follows, as per \eqref{zsd}:
\begin{align*}
\beta_{j,i}&=2\big(x_{j,i}-z_{j}(x_{j,i}-2)\big)/s_j(\boldsymbol{x}_j), 
 \quad\!\!\!\!\textup{for}\quad\!\!\!\!i,j=1,2\\
 \beta_{3,i}&=2\big(x_{1,i}-z_3(x_{1,i}-x_{2,i})\big)/s_3(\boldsymbol{x}), 
 \quad\!\!\!\!\textup{for}\quad\!\!\!\!i=1,2\\
 \beta_{3,i+2}&=2\big(x_{2,i}-z_3(x_{2,i}-x_{1,i})\big)/s_3(\boldsymbol{x}), 
 \quad\!\!\!\!\textup{for}\quad\!\!\!\!i=1,2. \\[-1cm]
\end{align*}

Since all eigenvalues of \(\boldsymbol{A}_\mathrm{d}\) in \eqref{mobile_ad} have zero real parts, by Lemma~3, any positive \(\gamma\) can be chosen; we set \(\gamma = 0.1\). Identical circular trajectories with differing frequencies for the robots lead to conflicts between tracking objective and collision avoidance. Prioritizing safety, the robots must deviate from their paths to avoid collisions. When conflicts arise, assume that we favor adjusting the second robot’s path to allow the first robot to maintain its trajectory. Thus, we select \(\boldsymbol{R} = \boldsymbol{I}_4\) and $\boldsymbol{Q}=\mathrm{\textbf{diag}}\big(\boldsymbol{Q}_1(\boldsymbol{e}_1),\boldsymbol{Q}_2(\boldsymbol{e}_2)\big)$, where $\boldsymbol{Q}_1(\boldsymbol{e}_1)=20\boldsymbol{I}_2/(\Vert \boldsymbol{e}_1\Vert^2 +0.01)$ $\boldsymbol{Q}_2(\boldsymbol{e}_2)=10\boldsymbol{I}_2/(\Vert \boldsymbol{e}_2\Vert^2 +0.01)\big)$, and $\boldsymbol{e}_j=\boldsymbol{x}_j-[v_{j,1},\,v_{j,2}]^\top$ for $j=1,2$. We set $q_{z_i}=10^{-3}$ for $i=1,2,3$. The \textsc{Matlab} simulation results are shown in Fig.~\ref{Fig_5}.

Fig.~\ref{Fig_5}(a), including its zoomed view, shows the robots tracking their circular paths in a 2D workspace while avoiding the obstacle. Robot~1 starts at \(\boldsymbol{x}_1(0) = [3,\,4]^\top\,\mathrm{m}\), while Robot~2 starts at \(\boldsymbol{x}_2(0) = [4,\,3]^\top\,\mathrm{m}\). Both successfully avoid the circular obstacle, dynamically adjusting their paths near the obstacle and each other to maintain safety constraints. Fig.~\ref{Fig_5}(b) shows bounded tracking errors, with Robot 1 achieving tighter tracking due to higher priority. Figs.~\ref{Fig_5}(c)--(e), especially their zoomed sections, confirms all safety constraints remain satisfied since \(s_1(\boldsymbol{x}_1)\), \(s_2(\boldsymbol{x}_2)\), and \(s_3(\boldsymbol{x})\) stay positive. The following points should be noted:

$\bullet$ Just before \(t = 4\pi\) seconds, the first robot completes its second full cycle while the slower second robot finishes its first. At this moment, the collision avoidance constraint nears its limit (see the blue solid  line in Fig.~\ref{Fig_5}(e)) as both robots occupy a similar workspace region, creating a potential conflict. To maintain the minimum distance, the controller lets the second robot temporarily deviate outward from its desired path, while the first robot continues with minimal deviation. This selective prioritization—enabled by the weighting matrices \(\boldsymbol{Q}_1(\boldsymbol{e}_1)\) and \(\boldsymbol{Q}_2(\boldsymbol{e}_2)\)—preserves tighter tracking for the first robot, with the second robot sacrificing some tracking performance to ensure safety.

$\bullet$ Shortly after \(t = 2\pi\) and \(t = 6\pi\) seconds, only Robot 1 nears the obstacle, while Robot 2 remains far from both the obstacle and the first robot (see the blue solid lines in Figs.~\ref{Fig_5}(d) and \ref{Fig_5}(e)). To ensure safety, the controller modifies the first robot’s path, temporarily increasing its tracking error (blue solid line in Fig.~\ref{Fig_5}(b)). Meanwhile, the second robot stays safely distant and tracks its path with nearly zero error (red dashed line in Fig.~\ref{Fig_5}(b)).

$\bullet$ Shortly after \(t = 4\pi\) seconds, both robots approach the obstacle region. To ensure obstacle avoidance, the controller adjusts the first robot’s path, causing a slight rise in its tracking error (blue solid line in Fig.~\ref{Fig_5}(b)). Meanwhile, the second robot, near both the obstacle and the first robot, must deviate further to satisfy safety constraints, resulting in a sharper increase in its tracking error (red dashed line in Fig.~\ref{Fig_5}(b)). Figs.~\ref{Fig_5}(d) and \ref{Fig_5}(e) show this tight scenario, with the second robot’s obstacle distance and inter-robot distance reaching their minimum safe values. Despite these challenges, all safety constraints remain satisfied.

For comparison, we simulate a safety-critical controller using the CBF-QP method. Here, the control input is \(\boldsymbol{u} = \bar{\boldsymbol{u}} + \delta \boldsymbol{u}\), where \(\bar{\boldsymbol{u}}\) achieves tracking and \(\delta \boldsymbol{u}\) enforces safety. We design \(\bar{\boldsymbol{u}} = [-\boldsymbol{e}_1^\top,\, -\boldsymbol{e}_2^\top]^\top + [2\cos(t),\, -2\sin(t),\, \cos(t),\, -\sin(t)]^\top\). The corrective term \(\delta \boldsymbol{u}\) is found by solving the following QP that minimally perturbs \(\bar{\boldsymbol{u}}\) while guaranteeing safety, as in \cite{ames2014control}.
\begin{align*}
\delta\boldsymbol{u}& = \operatorname*{arg\,min}_{\delta\boldsymbol{u} \in \mathbb{R}^m} \frac{1}{2} \| \delta\boldsymbol{u} \|^2 \tag{CBF-QP}\\
\text{s.t. } \dot{s}_j(\boldsymbol{x}_j) &\geq -\alpha_j \big(s_j(\boldsymbol{x}_j)\big), 
 \quad\!\!\!\!\textup{for}\quad\!\!\!\!j=1,2\\
 \dot{s}_3(\boldsymbol{x}) &\geq -\alpha_3 \big(s_3(\boldsymbol{x})\big) 
 \nonumber \\[-1cm]
\end{align*}

\(\alpha_1\), \(\alpha_2\), and \(\alpha_3\) are class \(\mathcal{K}\) functions, chosen as \(\alpha_1(s_j(\boldsymbol{x}_j))\! =\! 2 s_j(\boldsymbol{x}_j)\) for \(j\!=\!1,2\) and \(\alpha_3(s_3(\boldsymbol{x})) \!=\! 2 s_3(\boldsymbol{x})\). The resulting robot trajectories are shown in Fig.~\ref{Fig_5}(a), with the first robot in black dotted and the second in green dash-dotted lines. Tracking errors and safety constraints over time are illustrated in Figs.~\ref{Fig_5}(b)--(e). Defining $J=\int_0^{8\pi}[\boldsymbol{e}^\top\boldsymbol{Q}(\boldsymbol{e})\boldsymbol{e}+\boldsymbol{u}^\top\boldsymbol{u}]\mathrm{d}t$, 
where \(\boldsymbol{Q}(\boldsymbol{e})\) is as previously defined, yields \(J=334.47\) for the SSDRE controller and \(J=510.81\) for the CBF-QP controller. With a sampling rate of $100$ Hz, the simulation time for \(8\pi\) seconds was about $2.6$ seconds for SSDRE and $1.7$ seconds for CBF-QP. Although SSDRE takes slightly longer, it remains viable for real-time use and provides suboptimal solutions. 
\begin{remark}
Systems with two or more states admit infinitely many SDC parameterizations. According to \textup{\cite{huang1996nonlinear}}, under specific conditions, there exists an SDC representation enabling the SDRE controller to realize the optimal control law. While this result can be theoretically extended to the SSDRE method, finding such an SDC form is often extremely difficult or impossible \textup{\cite{ccimen2008state}}.
\end{remark}

\input{figures/simulation_05}

%% file: figures/simulation_01.tex
\begin{figure}
    \centering
    \begin{tikzpicture}[spy using outlines={square, magnification=2, size=1cm, connect spies}]
        \begin{scope}[xshift=0.5cm, yshift=0cm]
            \node[above] at (-1,4.5) {\scriptsize(a)};
            \begin{axis}[
                xmin=-3.5, xmax=3.5, 
                ymin=-3.25, ymax=3.25,
                xlabel={$x_1$},
                xlabel style={font=\footnotesize},
                ylabel style={rotate=-90},
                ylabel={$x_2$},
                ylabel style={font=\footnotesize},
                axis line style={-},
                grid=major, 
                xtick={-3,-2,-1,0,1,2,3}, 
                ytick={-3,-2,-1,0,1,2,3}, 
                    tick label style={font=\scriptsize},
                width=0.4\textwidth, 
                height=6.5cm,
                legend style={
                    font=\tiny,
                    draw=black,
                    fill=white,
                    legend pos=north west,
                    legend columns=1,
                    cells={anchor=west},   legend cell align=left,      inner xsep=0pt,
                    text depth=0pt,
                }]
                    \draw[fill=green!50, fill opacity=0.5, draw=green!50] (-3,-3) rectangle (3,3);              
                \addplot [blue, line width=1.5pt] table [x index=0, y index=1] {Fig_Data/Fig1_1.txt};
                \addlegendentry{SSDRE}
                \addplot [dotted,black, line width=1.5pt] table [x index=4, y index=5] {Fig_Data/Fig1_1.txt};
                \addlegendentry{SDRE}
                \addplot [dashdotted,orange, line width=1.5pt] table [x index=0, y index=1] {Fig_Data/Fig1_1_mpc.txt};
                \addlegendentry{NMPC}
                \addplot [dashed,red, thick] table [x index=2, y index=3] {Fig_Data/Fig1_1.txt};
                \addlegendentry{Desired}
                \addplot [
                    mark=square*,
                    mark options={
                        fill=blue, 
                        draw=black,
                        mark size=2pt
                    },
                    only marks
                ] coordinates {(2.5,-2.5)};
                \addplot [
                    mark=square*,
                    mark options={
                        fill=red, 
                        draw=red,
                        mark size=2pt
                    },
                    only marks
                ] coordinates {(2,2)};
            \end{axis}
        \end{scope}
                \begin{scope}[xshift=0cm, yshift=-3.5cm]
            \node[above] at (-1,2.) {\scriptsize (b)};
            \begin{axis}[
                xmin=0, xmax=10, 
                ymin=-37, ymax=20,
                xlabel={Time (s)},
                xlabel style={font=\footnotesize},
                ylabel style={rotate=-90},
                ylabel={$\boldsymbol{u}$},
                ylabel style={font=\footnotesize},
                axis line style={-},
                grid=major, 
                xtick={0,2,4,6,8,10}, 
                ytick={-30,-15,0,15},
                tick label style={font=\scriptsize},
                width=0.45\textwidth, 
                height=4cm,
                legend style={
                    font=\tiny,
                    draw=black,
                    fill=white,
                    legend pos=south east,
                    legend columns=1,
                    cells={anchor=west},   legend cell align=left,      inner xsep=0pt,
                    text depth=0pt,
                }]
                \addplot [blue, line width=1pt] table [x index=2, y index=0] {Fig_Data/Fig1_2.txt};
                \addlegendentry{$u_1$}
                \addplot [red, line width=1pt, dashed] table [x index=2, y index=1] {Fig_Data/Fig1_2.txt};
                \addlegendentry{$u_2$}
            \end{axis}
        \end{scope}
                        \begin{scope}[xshift=2.7cm, yshift=-3.15cm]
            \begin{axis}[
                xmin=0, xmax=0.5, 
                ymin=-38, ymax=20,
                axis line style={-},
                grid=none, 
                xtick={0,0.5,4,6,8,10},
                ytick={-20,0}, 
                width=3cm, 
                height=2.5cm,
    tick label style={font=\scriptsize},
]
                \addplot [blue, line width=1pt] table [x index=2, y index=0] {Fig_Data/Fig1_2.txt};
                \addplot [red, line width=1pt, dashed] table [x index=2, y index=1] {Fig_Data/Fig1_2.txt};
            \end{axis}
        \end{scope}
    \end{tikzpicture}
    \caption{\textsc{Matlab} simulation results for the non-conflicted case:
    (a) system trajectory under the proposed SSDRE controller compared with the conventional SDRE controller and a nonlinear MPC;
    (b) time evolution of the computed control signals using the proposed SSDRE controller.}\label{Fig_1}
    \vspace{-0.2cm}
\end{figure}
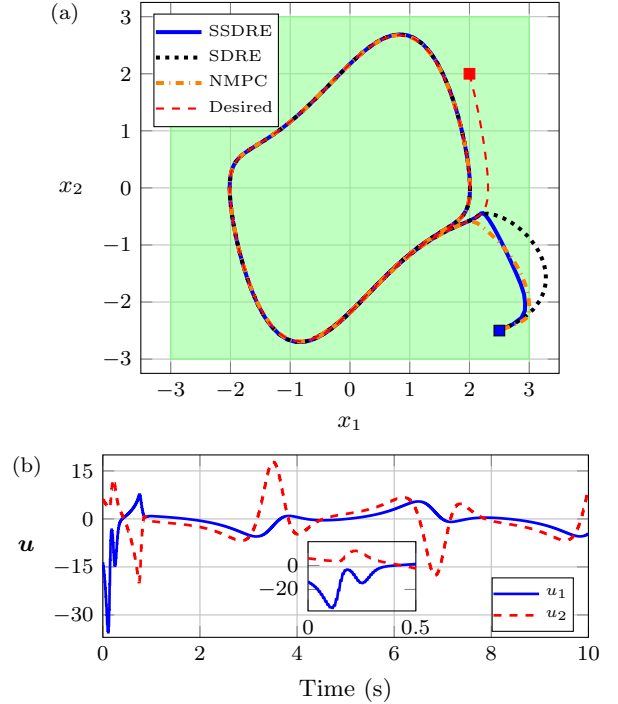

%% file: figures/simulation_02.tex
\begin{figure}
    \centering
    \begin{tikzpicture}[spy using outlines={square, magnification=2, size=1cm, connect spies}]
        \begin{scope}[xshift=0.5cm, yshift=0cm]
            \begin{axis}[
                xmin=-4, xmax=4, 
                ymin=-4, ymax=5.9,
                xlabel={$x_1$},
                xlabel style={font=\footnotesize},
                ylabel style={rotate=-90},
                ylabel={$x_2$},
                ylabel style={font=\footnotesize},
                axis line style={-},
                grid=major, 
                xtick={-4,-3,-2,-1,0,1,2,3,4}, 
                ytick={-4,-3,-2,-1,0,1,2,3,4},
                tick label style={font=\scriptsize},
                 width=0.4\textwidth, 
                height=6.5cm,
                legend style={
                    font=\tiny,
                    draw=black,
                    fill=white,
                    legend pos=north east,
                    legend columns=2,
                    cells={anchor=west},   legend cell align=left,      inner xsep=0pt,
                    text depth=0pt,
                }]
                \draw[fill=green!50, fill opacity=0.5, draw=green!50] (-3,-3) rectangle (3,3);
    
                \addplot [blue, line width=1.5pt] table [x index=0, y index=1] {Fig_Data/Fig2_1.txt};
                \addlegendentry{$q_z=10^2$}
                \addplot [dashdotted,orange, line width=1.5pt] table [x index=7, y index=8] {Fig_Data/Fig2_1.txt};
                \addlegendentry{$q_z=10^4$}
                \addplot [dotted,black, line width=1.5pt] table [x index=13, y index=14] {Fig_Data/Fig2_1.txt};
                \addlegendentry{SDRE}
                \addplot [dashed,red, thick] table [x index=2, y index=3] {Fig_Data/Fig2_1.txt};
                \addlegendentry{Desired}
                \addplot [
                    mark=square*,
                    mark options={
                        fill=blue, 
                        draw=black,
                        mark size=2pt
                    },
                    only marks
                ] coordinates {(1,1)};
                \addplot [
                    mark=square*,
                    mark options={
                        fill=red, 
                        draw=red,
                        mark size=2pt
                    },
                    only marks
                ] coordinates {(2,2)};
            \end{axis}
        \end{scope}
    \end{tikzpicture}
    \caption{\textsc{Matlab} simulation for the conflicted case (solution with a single barrier state): system trajectories with the SSDRE controller for $q_z=10^2$ and $q_z=10^4$ and the conventional SDRE controller.}\label{Fig_2}
    \vspace{-0.2cm}
\end{figure}
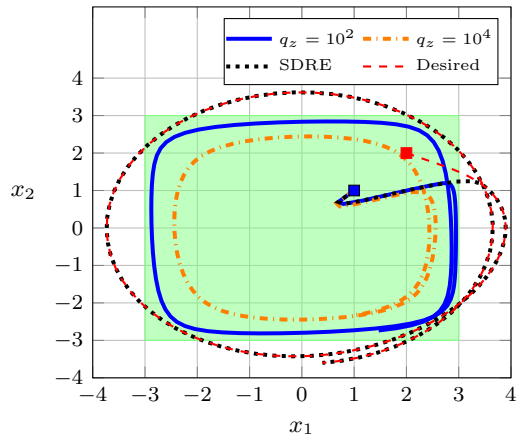

%% file: figures/simulation_03.tex
\begin{figure}
    \centering
    \begin{tikzpicture}[spy using outlines={square, magnification=2, size=1cm, connect spies}]
        \begin{scope}[xshift=0.5cm, yshift=0cm]
            \begin{axis}[
                xmin=-4, xmax=4, 
                ymin=-3.6, ymax=5.7,
                xlabel={$x_1$},
                xlabel style={font=\footnotesize},
                ylabel style={rotate=-90},
                ylabel={$x_2$},
                ylabel style={font=\footnotesize},
                axis line style={-},
                grid=major, 
                xtick={-4,-3,-2,-1,0,1,2,3,4}, 
                ytick={-3,-2,-1,0,1,2,3,4}, 
                tick label style={font=\scriptsize},
                width=0.4\textwidth, 
                height=6.5cm,
                legend style={
                    font=\tiny,
                    draw=black,
                    fill=white,
                    legend pos=north east,
                    legend columns=2,
                    cells={anchor=west},   legend cell align=left,      inner xsep=0pt,
                    text depth=0pt,
                }]
                  \draw[fill=green!50, fill opacity=0.5, draw=green!50] (-3,-3) rectangle (3,3);                              
                \addplot [blue, line width=1.5pt] table [x index=0, y index=1] {Fig_Data/Fig3.txt};
                \addlegendentry{$q_{z_1}\!=\!q_{z_2}\!=\!10$}
                
                \addplot [orange, dashdotted, line width=1.5pt] table [x index=2, y index=3] {Fig_Data/Fig3.txt};
                \addlegendentry{$q_{z_1}\!=\!10q_{z_2}\!=\!100$}
                 \addplot [red, thick, dashed] table [x index=6, y index=7] {Fig_Data/Fig3.txt};
                \addlegendentry{Desired}
                \addplot [black,dotted,line width=1.5pt] table [x index=4, y index=5] {Fig_Data/Fig3.txt};
                \addlegendentry{$10q_{z_1}\!=\!q_{z_2}\!=\!100$}

                \addplot [
                    mark=square*,
                    mark options={
                        fill=blue, 
                        draw=black,
                        mark size=2pt
                    },
                    only marks
                ] coordinates {(1,1)};
                \addplot [
                    mark=square*,
                    mark options={
                        fill=red, 
                        draw=red,
                        mark size=2pt
                    },
                    only marks
                ] coordinates {(2,2)};
            \end{axis}
        \end{scope}
    \end{tikzpicture}
    \caption{\textsc{Matlab} simulation results for conflicted case (solution with two barrier states): system trajectories with different barrier state weighting parameters.}\label{Fig_3}
    \vspace{-0.2cm}
\end{figure}
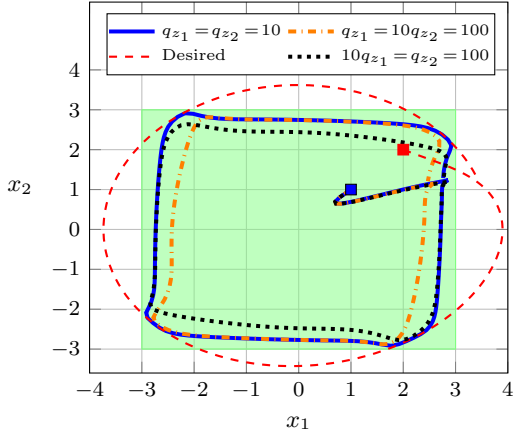

%% file: figures/simulation_04.tex
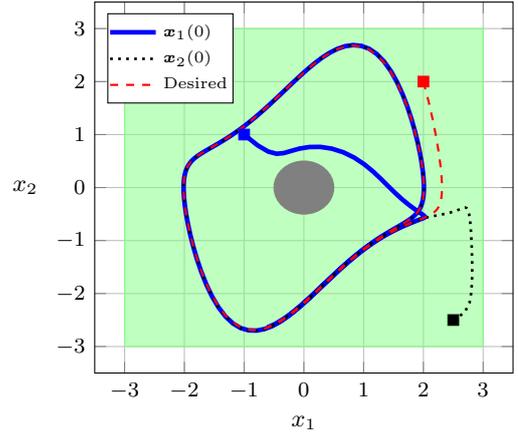
\begin{figure}
    \centering
    \begin{tikzpicture}[spy using outlines={square, magnification=2, size=1cm, connect spies}]
        \begin{scope}[xshift=0.5cm, yshift=0cm]
            \begin{axis}[
                xmin=-3.5, xmax=3.5, 
                ymin=-3.5, ymax=3.5,
                xlabel={$x_1$},
                xlabel style={font=\footnotesize},
                ylabel style={rotate=-90},
                ylabel={$x_2$},
                ylabel style={font=\footnotesize},
                axis line style={-},
                grid=major, 
                xtick={-3,-2,-1,0,1,2,3}, 
                ytick={-3,-2,-1,0,1,2,3},
                tick label style={font=\scriptsize},
                width=0.4\textwidth, 
                height=6.5cm,
                legend style={
                    font=\tiny,
                    draw=black,
                    fill=white,
                    legend pos=north west,
                    legend columns=1,
                    cells={anchor=west},   legend cell align=left,      inner xsep=0pt,
                    text depth=0pt,
                }]
                  \draw[fill=green!50, fill opacity=0.5, draw=green!50] (-3,-3) rectangle (3,3);                        \draw[fill=gray, fill opacity=1, draw=gray] (0,0) circle (0.5);                        
                \addplot [blue, line width=1.75pt] table [x index=0, y index=1] {Fig_Data/Fig4.txt};
                \addlegendentry{$\boldsymbol{x}_1(0)$}
                \addplot [black, dotted, line width=1pt] table [x index=4, y index=5] {Fig_Data/Fig4.txt};
                \addlegendentry{$\boldsymbol{x}_2(0)$}
                \addplot [red,  thick, dashed] table [x index=2, y index=3] {Fig_Data/Fig4.txt};
                \addlegendentry{Desired}
                \addplot [
                    mark=square*,
                    mark options={
                        fill=blue, 
                        draw=blue,
                        mark size=2pt
                    },
                    only marks
                ] coordinates {(-1,1)};
                \addplot [
                    mark=square*,
                    mark options={
                        fill=black, 
                        draw=black,
                        mark size=2pt
                    },
                    only marks
                ] coordinates {(2.5,-2.5)};
                \addplot [
                    mark=square*,
                    mark options={
                        fill=red, 
                        draw=red,
                        mark size=2pt
                    },
                    only marks
                ] coordinates {(2,2)};
            \end{axis}
        \end{scope}
    \end{tikzpicture}
    \caption{\textsc{Matlab} simulation results for non-conflicted case with three safety constraints.}\label{Fig_4}
    \vspace{-0.2cm}
\end{figure}

%% file: figures/simulation_05.tex
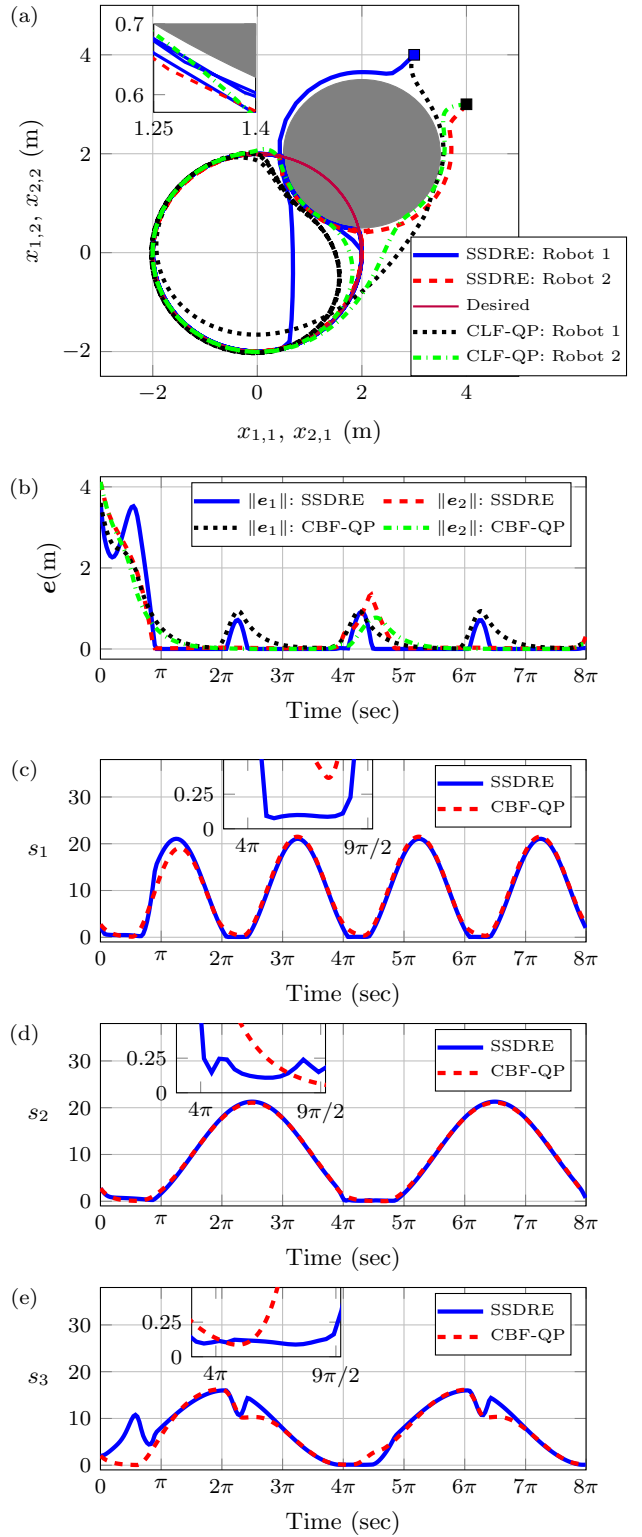
\begin{figure}
	\centering
	\begin{tikzpicture}[spy using outlines={square, magnification=2, size=1cm, connect spies}]

\begin{scope}[xshift=0.cm, yshift=0cm]
  \node[above] at (-1,4.5) {\scriptsize (a)};
  \begin{axis}[
      xmin=-3, xmax=5,
      ymin=-2.5, ymax=5,
      xlabel={$x_{1,1}$, $x_{2,1}$ (m)},
    xlabel style={font=\footnotesize},
      ylabel style={rotate=0},
      ylabel={$x_{1,2}$, $x_{2,2}$ (m)},
      ylabel style={font=\footnotesize},
      axis line style={-},
      grid=major,
      xtick={-2,0,2,4},
      ytick={-2,0,2,4},
      tick label style={font=\scriptsize},
      width=0.4\textwidth,
      height=6.5cm,
      legend style={
        font=\tiny,
        draw=black, fill=white,
        at={(1,0)}, anchor=south,
        legend columns=1,
        cells={anchor=west}, legend cell align=left,
        inner xsep=0pt}]
      \draw[fill=gray, fill opacity=1, draw=gray] (2,2) circle (1.5);

      \addplot [blue,           ultra thick]   table [x index=1, y index=2] {Fig_Data/Fig5.txt};
      \addlegendentry{SSDRE: Robot 1}

      \addplot [red, dashed,ultra thick] table [x index=3, y index=4] {Fig_Data/Fig5.txt};
      \addlegendentry{SSDRE: Robot 2}

      \addplot [purple, thick] table [x index=10, y index=11] {Fig_Data/Fig5.txt};
      \addlegendentry{Desired}

      \addplot [black,dotted,ultra thick]   table [x index=1, y index=2] {Fig_Data/Fig5_1.txt};
      \addlegendentry{CLF‑QP: Robot 1}

      \addplot [green, dashdotted, ultra thick] table [x index=3, y index=4] {Fig_Data/Fig5_1.txt};
      \addlegendentry{CLF‑QP: Robot 2}

      \addplot [mark=square*, mark options={fill=blue,  draw=black, mark size=2pt}, only marks]
               coordinates {(3,4)};
      \addplot [mark=square*, mark options={fill=black, draw=black, mark size=2pt}, only marks]
               coordinates {(4,3)};
  \end{axis}
\end{scope}

\begin{scope}[xshift=0.7cm, yshift=3.5cm]
  \begin{axis}[
      xmin=1.25, xmax=1.4,
      ymin=0.575,  ymax=0.7,
      axis line style={-,black},
      grid=none,
      xtick={1.25,1.4},
      ytick={0.6,0.7},
      tick label style={font=\scriptsize},
      width=0.165\textwidth,
      height=2.75cm]
      \draw[fill=gray, fill opacity=1, draw=gray] (2,2) circle (1.5);

      \addplot [blue,very thick]     table [x index=1, y index=2] {Fig_Data/Fig5.txt};
      \addplot [red, dashed, very thick]   table [x index=3, y index=4] {Fig_Data/Fig5.txt};

      \addplot [purple, thick]             table [x index=10, y index=11] {Fig_Data/Fig5.txt};

      \addplot [black,dotted, ultra thick]   table [x index=1, y index=2] {Fig_Data/Fig5_1.txt};
      \addplot [green, dashdotted, dashdotted, ultra thick] table [x index=3, y index=4] {Fig_Data/Fig5_1.txt};
  \end{axis}
\end{scope}

		\begin{scope}[xshift=0cm, yshift=-3.75cm]
			\node[above] at (-1,2.) {\scriptsize (b)};
			\begin{axis}[
				xmin=0, xmax=25.13, 
				ymin=-0.25, ymax=4.25,
				xlabel={Time (sec)},
                xlabel style={font=\footnotesize},
				ylabel style={rotate=0},
				ylabel={$\boldsymbol{e}$(m)},
                ylabel style={font=\footnotesize},
				axis line style={-},
				grid=major, 
				xtick={0, 3.1416, 6.2832, 9.4248, 12.5664, 15.7080, 18.8496, 22,25.1327},                    xticklabels={0, $\pi$, $2\pi$, $3\pi$, $4\pi$, $5\pi$, $6\pi$, $7\pi$,$8\pi$},
				ytick={0,2,4,6}, 
				tick label style={font=\scriptsize},
				width=0.45\textwidth, 
				height=4cm,
				legend style={
					font=\tiny,
					draw=black,
					fill=white,
					legend pos=north east,
					legend columns=2,
					cells={anchor=west},   legend cell align=left,      inner xsep=0pt,
					text depth=0pt,
				}]
				\addplot [blue, ultra thick] table [x index=0, y index=5] {Fig_Data/Fig5.txt};
				\addlegendentry{$\Vert \boldsymbol{e}_1\Vert$: SSDRE}
				\addplot [dashed,red, ultra thick] table [x index=0, y index=6] {Fig_Data/Fig5.txt};
				\addlegendentry{$\Vert \boldsymbol{e}_2\Vert$: SSDRE}
				\addplot [black,dotted, ultra thick] table [x index=0, y index=5] {Fig_Data/Fig5_1.txt};
				\addlegendentry{$\Vert \boldsymbol{e}_1\Vert$: CBF-QP}
				\addplot [dashdotted,green, ultra thick] table [x index=0, y index=6] {Fig_Data/Fig5_1.txt};
				\addlegendentry{$\Vert \boldsymbol{e}_2\Vert$: CBF-QP}
			\end{axis}
		\end{scope}
		         \begin{scope}[xshift=0cm, yshift=-7.5cm]
			\node[above] at (-1,2.) {\scriptsize (c)};
			\begin{axis}[
				xmin=0, xmax=25.13,
				ymin=-1, ymax=38,
				xlabel={Time (sec)},
                xlabel style={font=\footnotesize},
				ylabel style={rotate=-90},
				ylabel={$s_1$},
                ylabel style={font=\footnotesize},
				axis line style={-},
				grid=major,
				xtick={0, 3.1416, 6.2832, 9.4248, 12.5664, 15.7080, 18.8496, 22,25.1327},                    xticklabels={0, $\pi$, $2\pi$, $3\pi$, $4\pi$, $5\pi$, $6\pi$, $7\pi$,$8\pi$},
				ytick={0,10,20,30},
				tick label style={font=\scriptsize},
				width=0.45\textwidth,
				height=4cm,
				legend style={
					font=\tiny,
					draw=black,
					fill=white,
					legend pos=north east,
					legend columns=1,
					cells={anchor=west},   legend cell align=left,     inner xsep=0pt,
					text depth=0pt,
				}]
				\addplot [blue, ultra thick] table [x index=0, y index=7] {Fig_Data/Fig5.txt};
				\addlegendentry{SSDRE}
				\addplot [red, ultra thick,dashed] table [x index=0, y index=7] {Fig_Data/Fig5_1.txt};
				\addlegendentry{CBF-QP}
    			\end{axis}
		\end{scope}
		\begin{scope}[xshift=1.62cm, yshift=-6.cm]
			\begin{axis}[
				xmin=12.25, xmax=14.2,
				ymin=0, ymax=0.5,
				tick label style={font=\scriptsize, black},
				xtick={0, 3.1416, 6.2832, 9.4248, 12.5664, 14.137, 18.8496, 22,25.1327},            xticklabels={0, $\pi$, $2\pi$, $3\pi$, $4\pi$, $9\pi/2$, $6\pi$, $7\pi$,$8\pi$},
				ytick={0,0.25},
				axis line style={-,black},
				width=0.2\textwidth,
				height=2.5cm,
				]
				\addplot [blue, ultra thick] table [x index=0, y index=7] {Fig_Data/Fig5.txt};
				\addplot [red, ultra thick,dashed] table [x index=0, y index=7] {Fig_Data/Fig5_1.txt};
            \end{axis}
		\end{scope}
		\begin{scope}[xshift=0cm, yshift=-11cm]
			\node[above] at (-1,2.) {\scriptsize (d)};
			\begin{axis}[
				xmin=0, xmax=25.13,
				ymin=-1, ymax=38,
				xlabel={Time (sec)},
                xlabel style={font=\footnotesize},
				ylabel style={rotate=-90},
				ylabel={$s_2$},
                ylabel style={font=\footnotesize},
				axis line style={-},
				grid=major,
				xtick={0, 3.1416, 6.2832, 9.4248, 12.5664, 15.7080, 18.8496, 22,25.1327},                    xticklabels={0, $\pi$, $2\pi$, $3\pi$, $4\pi$, $5\pi$, $6\pi$, $7\pi$,$8\pi$},
				ytick={0,10,20,30},
				tick label style={font=\scriptsize},
				width=0.45\textwidth,
				height=4cm,
				legend style={
					font=\tiny,
					draw=black,
					fill=white,
					legend pos=north east,
					legend columns=1,
					cells={anchor=west},   legend cell align=left,     inner xsep=0pt,
					text depth=0pt,
				}]
				\addplot [blue, ultra thick] table [x index=0, y index=8] {Fig_Data/Fig5.txt};
				\addlegendentry{SSDRE}
				\addplot [red, ultra thick,dashed] table [x index=0, y index=8] {Fig_Data/Fig5_1.txt};
				\addlegendentry{CBF-QP}               
			\end{axis}
		\end{scope}
		\begin{scope}[xshift=1cm, yshift=-9.5cm]
			\begin{axis}[
				xmin=12.25, xmax=14.2,
				ymin=0, ymax=0.5,
				tick label style={font=\scriptsize, black},
				xtick={0, 3.1416, 6.2832, 9.4248, 12.5664, 14.137, 18.8496, 22,25.1327},                    xticklabels={0, $\pi$, $2\pi$, $3\pi$, $4\pi$, $9\pi/2$, $6\pi$, $7\pi$,$8\pi$},
				ytick={0,0.25},
				axis line style={-,black},
				width=0.2\textwidth,
				height=2.5cm,
				]
				\addplot [blue, ultra thick] table [x index=0, y index=8] {Fig_Data/Fig5.txt};
				\addplot [red, ultra thick,dashed] table [x index=0, y index=8] {Fig_Data/Fig5_1.txt};
            \end{axis}
		\end{scope}
		\begin{scope}[xshift=0cm, yshift=-14.5cm]
	\node[above] at (-1,2.) {\scriptsize (e)};
	\begin{axis}[
		xmin=0, xmax=25.13,
		ymin=-1, ymax=38,
		xlabel={Time (sec)},
        xlabel style={font=\footnotesize},
		ylabel style={rotate=-90},
		ylabel={$s_3$},
        ylabel style={font=\footnotesize},
		axis line style={-},
		grid=major,
		xtick={0, 3.1416, 6.2832, 9.4248, 12.5664, 15.7080, 18.8496, 22,25.1327},  xticklabels={0, $\pi$, $2\pi$, $3\pi$, $4\pi$, $5\pi$, $6\pi$, $7\pi$,$8\pi$},
		ytick={0,10,20,30},
		tick label style={font=\scriptsize},
		width=0.45\textwidth,
		height=4cm,
		legend style={
			font=\tiny,
			draw=black,
			fill=white,
			legend pos=north east,
			legend columns=1,
			cells={anchor=west},   legend cell align=left,     inner xsep=0pt,
			text depth=0pt,
		}]
		\addplot [blue, ultra thick] table [x index=0, y index=9] {Fig_Data/Fig5.txt};
		\addlegendentry{SSDRE}
		\addplot [red, ultra thick,dashed] table [x index=0, y index=9] {Fig_Data/Fig5_1.txt};
		\addlegendentry{CBF-QP}
	\end{axis}
\end{scope}
\begin{scope}[xshift=1.2cm, yshift=-13cm]
	\begin{axis}[
		xmin=12.25, xmax=14.2,
		ymin=0, ymax=0.5,
		tick label style={font=\scriptsize, black},
		xtick={0, 3.1416, 6.2832, 9.4248, 12.5664, 14.137, 18.8496, 22,25.1327},                    xticklabels={0, $\pi$, $2\pi$, $3\pi$, $4\pi$, $9\pi/2$, $6\pi$, $7\pi$,$8\pi$},
		ytick={0,0.25},
		axis line style={-,black},
		width=0.2\textwidth,
		height=2.5cm,
		]
		\addplot [blue, ultra thick] table [x index=0, y index=9] {Fig_Data/Fig5.txt};
		\addplot [red, ultra thick,dashed] table [x index=0, y index=9] {Fig_Data/Fig5_1.txt};
	\end{axis}
\end{scope}
	\end{tikzpicture}
	\caption{\textsc{Matlab} simulation results for mobile robots collision avoidance  using the SSDRE and CBF-QP controllers: (a) the trajectories of the robots, (b) time evolution of the tracking errors, (c)--(e) time evolution of the safety constraints.}\label{Fig_5}
	\vspace{-0.5cm}
\end{figure}

%% file: Experiment.tex
\section{Experimental Results}
\label{sec:experiment}
For the implementation section, we consider a cable-suspended planar parallel robot (CSPPR). A schematic diagram of this robots is shown in Fig.~\ref{fig:cable_robot_schematic}.
The nonlinear dynamics of the system are described as follows  \cite{omidi2023time}:
\begin{align}\label{eq:cable_robot_dynamic}
\begin{bmatrix}
  \ddot{q}_1 \\[6pt]
  \ddot{q}_2
\end{bmatrix}
=
\frac{1}{m_r}
\begin{bmatrix}
  \dfrac{w - q_\mathrm{1}}{l_1}  & -\dfrac{w + q_\mathrm{1}}{l_2} \\[8pt]
   \dfrac{h - q_\mathrm{2}}{l_1} & \dfrac{h - q_\mathrm{2}}{l_2}
\end{bmatrix}
\begin{bmatrix}
  T_1 \\[6pt]
  T_2
\end{bmatrix}
\;-\;
\begin{bmatrix}
  0 \\
  g_r
\end{bmatrix},
\end{align}
where $2w$ is the horizontal distance between the two tangent points of the pulleys; $h$ is the vertical distance from the tangent point to the origin; $m_r$ is the mass of the end effector; and $g_r=9.81$m/sec$^2$ is the acceleration due to gravity. The positions in the $X$ and $Y$ directions are given by $q_\mathrm{1}$ and $q_\mathrm{2}$, respectively. The corresponding accelerations are represented by $\ddot{q}_\mathrm{1}$ and $\ddot{q}_\mathrm{2}$. 
The tension in the cables is considered as the input vector, where \(T_1\) and \(T_2\) represent the tensions in the right and left cables, respectively. The length of cables are given \cite{omidi2023time}:
\begin{align*}
l_{1}
    &= \sqrt{\bigl(w - q_{1}\bigr)^{2} + \bigl(h - q_{2}\bigr)^{2}},\\[4pt]
l_{2}
    &= \sqrt{\bigl(w + q_{1}\bigr)^{2} + \bigl(h - q_{2}\bigr)^{2}}.
\end{align*}
As illustrated in Fig.~\ref{fig:cable_robot_schematic}, the
CSPPR is driven by two $12\,$V DC gearmotors (500 RPM) equipped with magnetic encoders that
rotate $2\,$cm‑radius pulleys.  
The motors are powered through a HiLetgo~TB6612 dual H‑bridge shield mounted on an \textit{Arduino Uno}, while an \textit{Arduino Mega 2560} acquires the quadrature encoder pulses and streams state data to the host computer at approximately $20\,$Hz.
The host is a personal laptop running \textsc{Matlab} R2024a on a
12th‑Gen Intel\textsuperscript{\textregistered} Core\textsuperscript{\texttrademark}
i7‑12700H CPU (2.30GHz, 16GB RAM).
Inverse‑kinematics relations convert pulley rotations into cable lengths and, by differentiation, cable velocities; from these signals the Cartesian position and velocity of the end‑effector are achieved in real time. The robot parameters are \( m_r = 0.2 \, \text{Kg} \), \( h = 0.4 \, \text{m} \), and \(w= 0.9 \, \text{m} \). Letting $\boldsymbol{x} \coloneqq [q_1,\, \dot{q}_1,\, q_2,\, \dot{q}_2]^\top$ and $\boldsymbol{u} \coloneqq [T_1,\, T_2]^\top$, the objective is to design a controller that ensures: i) the end effector avoids entering a circular region centered at $(-0.2, 0)$~m with a radius of $0.1$~m, and ii) the end effector reaches the target position at $[v_1,\,v_2]=[-0.35,\, 0.05]$~m. To represent the safety constraint, we define $
\mathcal{S}=\{ \boldsymbol{x} \in \mathbb{R}^4 \ | \ s(x_1, x_3) \coloneqq (x_1 + 0.2)^2 + x_3^2 - 0.1^2 > 0 \}$. As per \eqref{asdc1} and \eqref{asdcv1}, we have $\boldsymbol{A}_\mathrm{d}=\boldsymbol{0}$ and $\boldsymbol{H}_\mathrm{d}=\boldsymbol{I}_{2}$. Since $\lambda(\boldsymbol{A}_\mathrm{d})=\{0,0\}$, we set $\gamma=0.01$. Since the safety and tracking objectives are not in conflict, the barrier state is given by $z = \left(e_1^2 + e_3^2\right) / s(x_1, x_3)$, with $e_1 = x_1 + 0.35$ and $e_3 = x_3 - 0.05$ (as noted in Remark~\ref{remark_new_z}). 
Following \eqref{sdcxdd}, we assign $\boldsymbol{\beta} = \boldsymbol{\alpha}_\mathrm{d} = \boldsymbol{0}$ and define $\boldsymbol{\alpha} = [0,\, \alpha_2,\, 0,\, \alpha_4]^\top$, with
\begin{align}\label{ggt1}
\begin{split}
\alpha_2 &= \frac{2(x_1 + 0.35) - 2z(x_1 + 0.2)}{s}, \\
\alpha_4 &= \frac{2(x_3 - 0.05) - 2zx_3}{s}. \\[-0.5cm]
\end{split}
\end{align}

To accommodate the presence of $g_r$ in \eqref{eq:cable_robot_dynamic}, $\boldsymbol{A}_z = [a_{i,j}]$ and $\boldsymbol{G}_z = [g_{i,k}]$ are defined for $i,j=1{:}7$ and $k=1,2$, with their non-zero elements specified as follows: the diagonal entries of $\boldsymbol{A}_z$ are set to $-\gamma$, while $a_{1,2}$ and $a_{3,4}$ are assigned the value $1$. The elements $a_{4,5}$ and $a_{4,6}$ account for the influence of $g_r$, defined as $-g_r v_1/(v_1^2 + v_2^2)$ and $-g_r v_2/(v_1^2 + v_2^2)$, respectively. Additionally, $a_{7,2}$ and $a_{7,4}$ correspond to $\alpha_2$ and $\alpha_4$ in \eqref{ggt1}. For the input matrix $\boldsymbol{G}_z$, the non-zero entries include $g_{2,1}=(w-x_1)/(m_rl_1)$, $g_{2,2}=-(w+x_1)/(m_rl_2)$, $g_{4,1}=(h-x_3)/(m_rl_1)$, and $g_{4,2}=(h-x_3)/(m_rl_2)$. We set the weighting matrices in the cost function \eqref{jjz} as $\boldsymbol{Q}=\boldsymbol{I}_2/\big(\big\Vert[e_1,e_3]\big\Vert^2+10^{-3}\big)$, $\boldsymbol{R}=100\boldsymbol{I}_2$, and $q_z=10^{-2}/(1+z)$. For $\boldsymbol{x}_0 = [-0.02,\,0,\,0,\,0]^\top$, the system trajectories obtained under the  SSDRE and SDRE controllers are presented in Fig.~\ref{Fig_6}. In this figure, the blue solid line represents the system trajectory obtained from the experimental setup using the SSDRE controller, the red dashed line corresponds to the simulated trajectory in \textsc{Matlab} under the SSDRE controller, and the green dotted line shows the experimental trajectory obtained with the SDRE controller. The total duration of each experiment and simulation is $10$ seconds. As illustrated in Fig.~\ref{Fig_6} and demonstrated in the recorded video\footnote{ link: \textcolor{blue}{\url{https://www.youtube.com/watch?v=l6PA_QMXjpU}}}, the SSDRE controller successfully enforces the safety constraint while achieving the tracking objective in both simulation and experiment. In contrast, the SDRE controller is only capable of satisfying the tracking objective and fails to satisfy the safety constraint. In the experimental tests, the end-effector required $8.34$ seconds to reach and stay within the bound $\big\Vert[e_1,e_3]\big\Vert < 0.01$ when using the SSDRE controller, compared to $6.71$ seconds for the SDRE controller. The SSDRE controller’s $24\%$ longer convergence time reflects its prioritized constraint: a collision-avoidance trajectory that sacrifices speed for safety. By defining the performance metric \( J_e = \int_0^{10} \big\Vert [e_1,\, e_3] \big\Vert \, \mathrm{d}t \), the resulting values are \(0.76\)~m$\cdot$sec for the SSDRE controller and \(0.43\)~m$\cdot$sec for the SDRE controller.

\input{figures/experiment_01}
\input{figures/Experiment_PP}

%% file: figures/experiment_01.tex
\begin{figure}[htbp]
  \centering
  \begin{tikzpicture}[every node/.style={text=white}]
    ----------------------------------------------------------------
    \node[anchor=south west, inner sep=0] (image) at (0,0)
      {\includegraphics[width=0.9\linewidth]{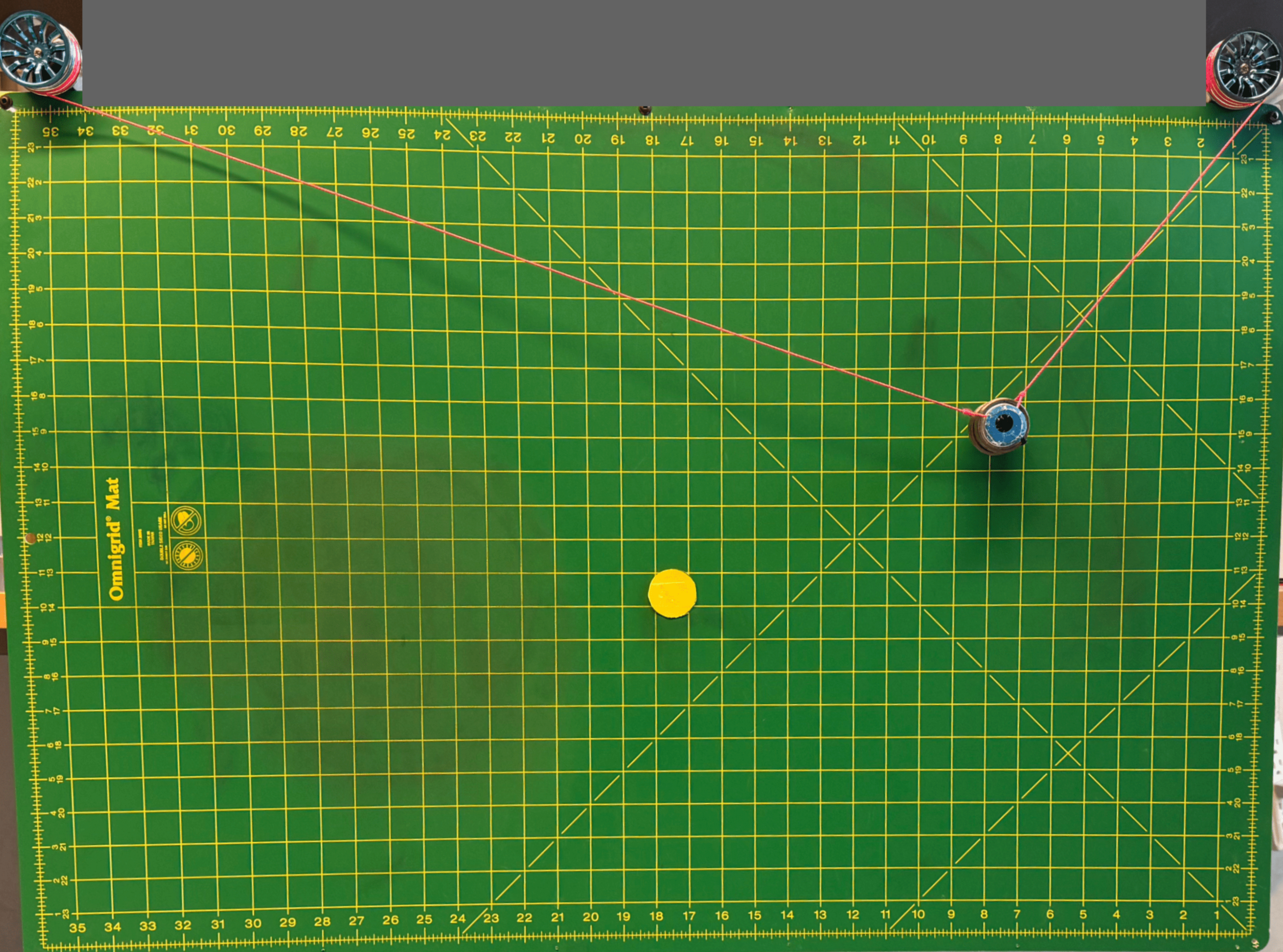}}; 
    ----------------------------------------------------------------
    \begin{scope}[x={(image.south east)}, y={(image.north west)}]

      \node at (0.20,0.95) {Left Pulley};
      \node at (0.80,0.95) {Right Pulley};

         \draw[-,line width=1.5,white]
            (0.785,0.555) -- (0.98,0.88)
    node[] {};

      \draw[-,line 
       width=1.5,white]
      (0.785,0.55) -- 
      (0.03,0.9)
      node[] {};

      \draw[->,>=latex,line width=2,red]
        (0.785,0.555) -- (0.785,0.4)
        node[pos=1,yshift=-2pt] {$m_rg_r$};

      \draw[->,>=latex,line width=2,red]
        (0.785,0.555) -- (0.875,0.71)
        node[pos=0.5,xshift=-3pt,yshift=10pt] {$u_{1}$};

      \draw[->,>=latex,line width=2,red]
        (0.78,0.555) -- (0.66,0.61)
        node[pos=0.5,xshift=-10pt,yshift=-5pt] {$u_{2}$};

      \draw[<->,>=latex,line width=1.5]
        (0.03,0.89) -- (0.98,0.87)
        node[pos=0.5,yshift=-5pt] {$2w$};

      \draw[<->,>=latex,line width=1.5]
        (0.03,0.9) -- (0.04,0.35)
        node[pos=0.5,xshift=5pt,yshift=0pt] {$h$};

      \draw[dashed,line width=1.5]
        (0.06,0.36) -- (0.53,0.365);

      \coordinate (origin) at (0.52,0.365);
      \fill (origin) circle[radius=1pt];
      \draw[->,>=latex,line width=2] (origin) -- ++(0.20,0) node[below] {$X$};
      \draw[->,>=latex,line width=2] (origin) -- ++(0,0.20) node[left] {$Y$};
      \node at (0.50,0.32) {$O$};

      \node at (0.9,0.55) {\tiny End Effector};

      \node at (0.25,0.75) {$l_{2}$};
      \node at (0.88,0.80) {$l_{1}$};
    \end{scope}
  \end{tikzpicture}

  \caption{Schematic of the CSPPR experimental setup:
  the right and left pulleys (top corners) are separated by
  $2w$ horizontally and by a vertical offset~$h$ from the origin~$O$;
  cable lengths $l_{1}(t)$ and $l_{2}(t)$ are actuated by input
  tensions $u_{1}$ and~$u_{2}$ (red arrows); the end‑effector mass
  experiences gravity $m_rg_r$; inertial $X$--$Y$ axes are
  shown for reference.}
  \label{fig:cable_robot_schematic}
\end{figure}
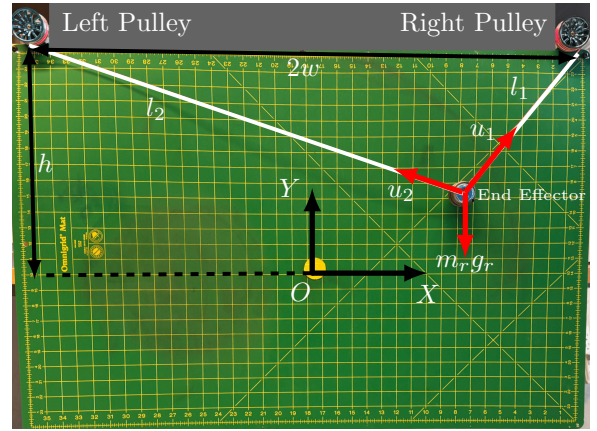

%% file: figures/Experiment_PP.tex
\begin{figure}[htbp]
    \centering
    \begin{tikzpicture}[
        spy using outlines={
            square,
            magnification=2,
            size=1cm,
            connect spies
        }]
        \begin{axis}[
            xmin=-0.375, xmax=0,
            ymin=-0.195, ymax=0.145,
            xlabel={$x_1$ (m)},
            xlabel style={font=\footnotesize},
            ylabel={$x_3$ (m)},
            ylabel style={font=\footnotesize},
            ylabel style={rotate=0},
            axis line style={-},
            grid=major,
            xtick={-0.35,-0.2,0},
            ytick={-0.1,0,0.1},
		tick label style={font=\scriptsize},
            width=0.4\textwidth,
            height=6.5cm,
            legend style={
                font=\tiny,
                draw=black,
                fill=white,
                legend pos=south east,
                cells={anchor=west},
                inner xsep=0pt,
                text depth=0pt,
            },
        ]
            \draw[fill=gray, draw=gray] (-0.2,0) circle (0.1);

            \addplot[blue, line width=1.5pt]
                table[x index=2, y index=3] {Fig_Data/Experiment.txt};
            \addlegendentry{SSDRE: experiment}

            \addplot[dotted, green, line width=1.5pt]
                table[x index=0, y index=1] {Fig_Data/Experiment.txt};
            \addlegendentry{SDRE: experiment}

            \addplot[red, dashed, line width=1.5pt]
                table[x index=0, y index=1] {Fig_Data/cable_sim.txt};
            \addlegendentry{SSDRE: simulation}

            \addplot[
                only marks,
                mark=square*,
                mark options={fill=blue, draw=blue, mark size=2pt},
            ] coordinates {(-0.022,0.001)};

            \addplot[
                only marks,
                mark=diamond*,
                mark options={fill=red, draw=red, mark size=4pt},
            ] coordinates {(-0.35,0.05)};
        \end{axis}
    \end{tikzpicture}
\caption{Trajectories of the CSPPR under the SSDRE and SDRE controllers. A demonstration video is available at: \textcolor{blue}{\url{https://www.youtube.com/watch?v=l6PA_QMXjpU}}}\label{Fig_6}
\end{figure}
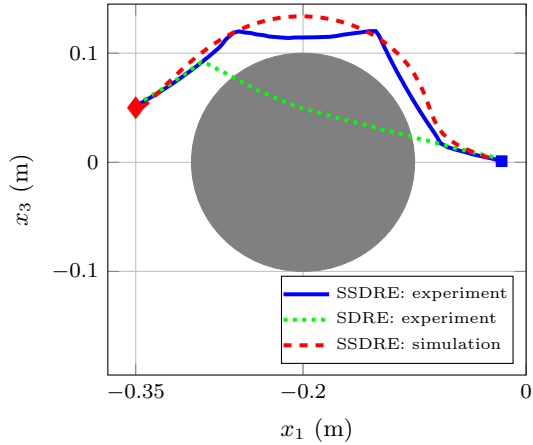